\documentclass[aps,prl,reprint,amsmath,amssymb,superscriptaddress,showpacs]{revtex4-2}
\usepackage{graphicx}
\usepackage{dcolumn}
\usepackage{bm}
\usepackage[colorlinks,linkcolor=blue,anchorcolor=blue,citecolor=blue,urlcolor=blue,filecolor=blue,menucolor=blue,runcolor=blue]{hyperref}

\begin{document}
\title{Anisotropic ${\mathbf{\textit{c}}}\ensuremath{-}{\mathbf{\textit{f}}}$ hybridization in the ferromagnetic quantum critical metal ${\mathbf{CeRh}}_{6}{\mathbf{Ge}}_{4}$}
\author{Yi Wu}
\affiliation{Center for Correlated Matter and Department of Physics, Zhejiang University, Hangzhou 310058, China}
\author{Yongjun Zhang}
\affiliation{Center for Correlated Matter and Department of Physics, Zhejiang University, Hangzhou 310058, China}
\affiliation{Institute for Advanced Materials, Hubei Normal University, Huangshi 435002, China}
\author{Feng Du}
\affiliation{Center for Correlated Matter and Department of Physics, Zhejiang University, Hangzhou 310058, China}
\author{Bin Shen}
\affiliation{Center for Correlated Matter and Department of Physics, Zhejiang University, Hangzhou 310058, China}
\author{Hao Zheng}
\affiliation{Center for Correlated Matter and Department of Physics, Zhejiang University, Hangzhou 310058, China}
\author{Yuan Fang}
\affiliation{Center for Correlated Matter and Department of Physics, Zhejiang University, Hangzhou 310058, China}
\author{ Michael Smidman}
\affiliation{Center for Correlated Matter and Department of Physics, Zhejiang University, Hangzhou 310058, China}
\author{ Chao Cao}
\affiliation{Department of Physics, Hangzhou Normal University, Hangzhou, China}
\author{Frank Steglich}
\affiliation{Center for Correlated Matter and Department of Physics, Zhejiang University, Hangzhou 310058, China}
\affiliation{Max Planck Institute for Chemical Physics of Solids, 01187 Dresden, Germany}
\author{Huiqiu Yuan}
\email {hqyuan@zju.edu.cn}
\affiliation{Center for Correlated Matter and Department of Physics, Zhejiang University, Hangzhou 310058, China}
\affiliation{Zhejiang Province Key Laboratory of Quantum Technology and Device, Zhejiang University, Hangzhou, China}
\affiliation{State Key Laboratory of Silicon Materials, Zhejiang University, Hangzhou 310058, China}
\author{Jonathan D. Denlinger}
\affiliation{Advanced Light Source, E.O. Lawrence Berkeley National Lab, Berkeley, CA 94720, USA}
\author{Yang Liu}
\email {yangliuphys@zju.edu.cn}
\affiliation{Center for Correlated Matter and Department of Physics, Zhejiang University, Hangzhou 310058, China}
\affiliation{Zhejiang Province Key Laboratory of Quantum Technology and Device, Zhejiang University, Hangzhou, China}
\date{\today}%
\addcontentsline{toc}{chapter}{Abstract}

\begin{abstract}
  Heavy fermion compounds exhibiting a ferromagnetic quantum critical point have attracted considerable interest. Common to two known cases, i.e., CeRh$_6$Ge$_4$ and YbNi$_4$P$_2$, is that the $4f$ moments reside along chains with a large inter-chain distance, exhibiting strong magnetic anisotropy that was proposed to be vital for the ferromagnetic quantum criticality. Here we report an angle-resolved photoemission study on CeRh$_6$Ge$_4$, where we observe sharp momentum-dependent $4f$ bands and clear bending of the conduction bands near the Fermi level, indicating considerable hybridization between conduction and $4f$ electrons. The extracted hybridization strength is anisotropic in momentum space and is obviously stronger along the Ce chain direction.The hybridized $4f$ bands persist up to high temperatures, and the evolution of their intensity shows clear band dependence. Our results provide spectroscopic evidence for anisotropic hybridization between conduction and $4f$ electrons in CeRh$_6$Ge$_4$, which could be important for understanding the electronic origin of the ferromagnetic quantum criticality.
\end{abstract}

\maketitle

Recently, a ferromagnetic (FM) quantum critical point (QCP) and associated strange metal behavior has been discovered in CeRh$_6$Ge$_4$ \cite{shen2020strange}. While most known QCPs in heavy fermion (HF) metals are of the antiferromagnetic (AFM) type \cite{gegenwart2008quantum}, the FM QCPs were thought to be prohibited in clean itinerant FM systems due to the influence of long-range correlation effects \cite{Belitz1999first,brando2016metallic,Chubukov2004instability}. The observation of a FM QCP in pressurized pristine CeRh$_6$Ge$_4$ opens up new opportunities to understand quantum critical phenomena and to unravel the origin of the strange metallic behavior \cite{RevModPhys.73.797,Lohneysen2007Fermi}. Common to two $4f$-electron HF systems exhibiting a FM QCP, i.e., pressurized CeRh$_6$Ge$_4$  \cite{shen2020strange} and As-substituted YbNi$_4$P$_2$ \cite{steppke2013ferromagnetic}, is that the $4f$ moments reside along chains, with large inter-chain distances and much smaller spacing along the chain \cite{shen2020strange,krellner2011ferromagnetic}. Such a chain-like configuration could lead to dominant magnetic exchange interactions along the chain direction, which was theoretically proposed to be key for the observed FM QCP \cite{shen2020strange}. Since the magnetic exchange interaction is electronic in nature, the dispersion of quasiparticle bands could also be highly anisotropic, possibly leading to quasi-one-dimensional (1D) electronic states as proposed in YbNi$_4$P$_2$ \cite{krellner2011ferromagnetic}. Quasi-1D chains of $4f$ moments are also reported in a few other HF systems, e.g.,  in CeCo$_2$Ga$_8$ where non-Fermi-liquid phenomena were also observed \cite{wang2017heavy,Cheng2019PRM}. However, whether such a chain-like arrangement of $4f$ moments may indeed lead to quasi-1D or anisotropic $4f$ bands has not yet been verified by momentum-resolved measurements, such as angle-resolved photoemission spectroscopy (ARPES). In the spin-triplet superconductor candidate UTe$_2$ with possible ferromagnetic fluctuations, a recent ARPES study indeed revealed the quasi-1D conduction bands resulting from the U and Te chains \cite{Miao2020PRL}.

\begin{figure}[ht]
\includegraphics[width=1.\columnwidth]{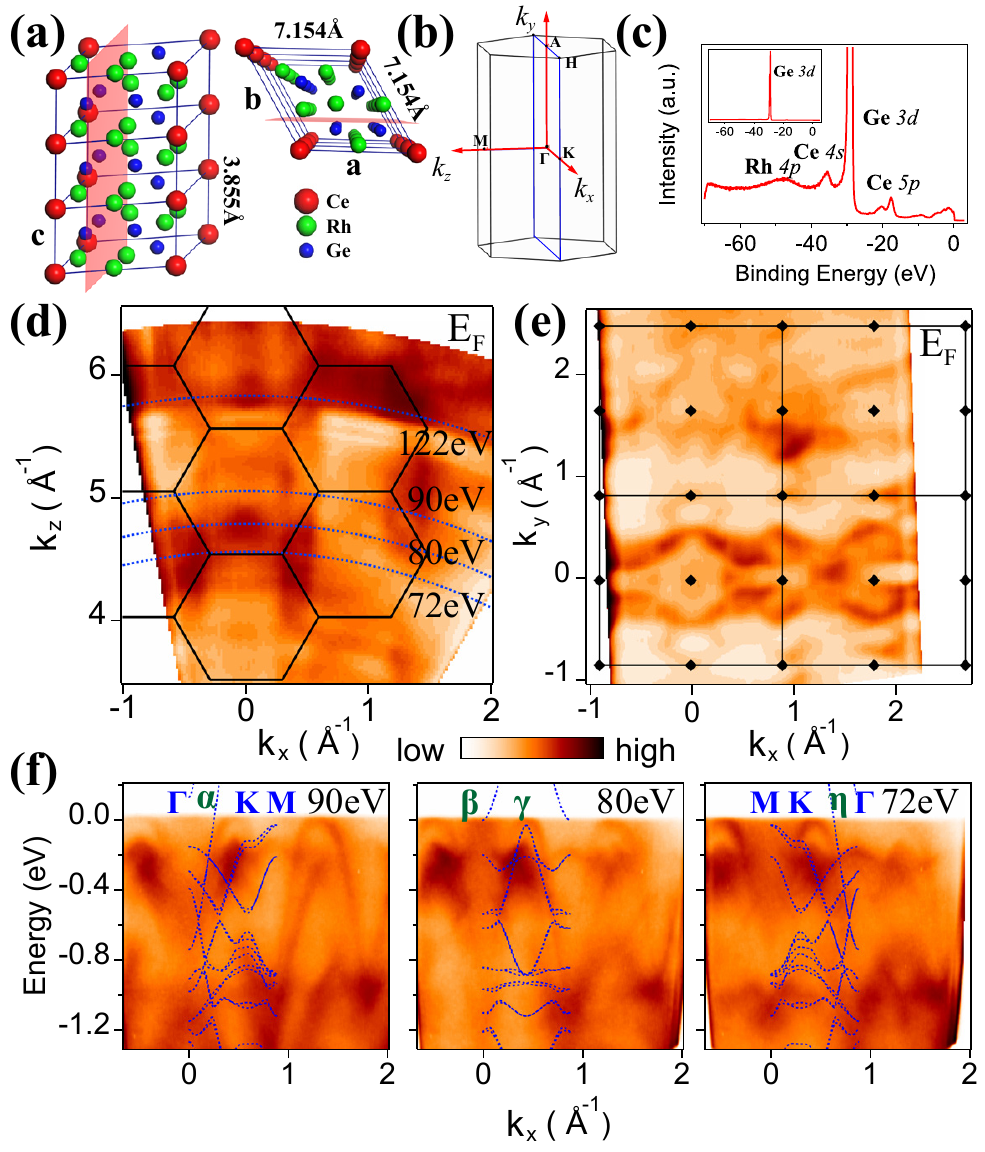}
\centering
\caption{(Color online). (a) Crystal structure of CeRh$_6$Ge$_4$. The light red plane indicates the possible cleavage surface. (b) Bulk Brillouin zone (BZ) and the momentum axes defined in this paper. $k_z$ is defined along the bulk [010] direction. (c) Large-range scan showing core levels (inset is a full-scale view). (d) Photon-energy dependent scan (30-150 eV) along the in-plane $k_x$ direction at $E_F$. An inner potential of $\sim$12 eV is used for the conversion to $k_z$. The data is plotted using a color code legend shown at the bottom. (e) $k_x-k_y$ FS map at 90 eV ($k_z\sim$ 0). The black lines (dots) in (d,e) indicate the bulk BZ boundaries (high symmetry points). (f) Energy-momentum dispersions at representative $k_z$ cuts, together with the localized $4f$ calculation shown in the first BZ (dashed blue curves). High symmetry momenta points ($\Gamma$, K, M) and bands crossing $E_F$ ($\alpha$, $\beta$, $\gamma$, $\eta$) are labeled.
}
\label{Fig1}
\end{figure}

Here we present ARPES results on CeRh$_6$Ge$_4$, a FM HF compound with Curie temperature $T_C$ = 2.5 K at ambient pressure. $T_C$ can be continuously suppressed to zero by pressure, resulting in a FM QCP at $p_c$ = 0.8 GPa \cite{shen2020strange,kotegawa2019indication}. It was theoretically proposed that the FM QCP in the zero temperature limit involves a simultaneous breakdown of the Kondo screening, resulting in an abrupt jump in Fermi surface (FS) volume from a 'large FS' incorporating $4f$ electrons in the high-pressure paramagnetic phase to a 'small FS' that does not contain $4f$ electrons in the low-pressure FM phase \cite{shen2020strange,cao2020pressure}. On the other hand, the small ordered moment in the FM state ($\sim$0.28 $\mu_B$/Ce), the small magnetic entropy released at $T_C$ ($\sim$0.19 R$\ln$2) and the large specific-heat coefficient extracted from above $T_C$ ($\sim$0.25 J$\cdot$mol$^{-1}\cdot$K$^{-2}$) imply that Kondo screening operates dynamically here \cite{matsuoka2015ferromagnetic,hu2020quantum}. The magnetic part of the resistivity shows a characteristic hump at $\sim$80 K \cite{supplementary}, suggesting Kondo screening well above $T_C$ (likely involving excited crystal electric field (CEF) states). These results therefore call for spectroscopic measurements to understand the local or itinerant nature of the Ce $4f$ electrons.

CeRh$_6$Ge$_4$ crystallizes in a simple hexagonal structure (Fig.~\ref{Fig1}(a)), with Ce aligning in chains along the $c$ axis with an intra-chain Ce-Ce distance of 3.855 {\AA} and an inter-chain separation of 7.154 {\AA} in the $a-b$ plane. It can be cleaved along the (010) surface, with the Ce chains lying in-plane (along $k_y$ defined in Fig.~\ref{Fig1}(b)) \cite{supplementary}. Large-range energy scans reveal core levels from Rh $4p$, Ce $4s$, Ce $5p$ and Ge $3d$ electrons, with the dominant contribution from Ge $3d$. This implies that the surface is likely Ge-terminated, supporting the bulk character of the measured $4f$ spectra. Photon-energy dependent scans (Fig.~\ref{Fig1}(d)) reveal periodic structures in accordance with the expected bulk BZs, despite large variation in the photoemission cross section, particularly near the Ce resonance edge (122 eV). The band periodicity can be better visualized from the $k_x-k_y$ map in Fig.~\ref{Fig1}(e), where wiggling bands near $k_y\sim \pm$0.3 {\AA}$^{-1}$ extending along $k_x$ can be observed, with a periodicity consistent with the bulk BZs. In Fig.~\ref{Fig1}(f), three photon-energy cuts are presented, corresponding to $k_z\sim$ 0, 0.25~$b^*$ and 0.5~$b^*$ ($b^*$ is the reciprocal lattice vector corresponding to $b$ in Fig.~\ref{Fig1}(a)). This $k_z$ conversion is based on a detailed comparison of the experimental data with density-functional theory (DFT) calculations treating $4f$ electrons as core electrons, i.e., the 'localized $4f$' calculation \cite{supplementary}, which yields an estimated inner potential of $\sim$12 eV. Experimentally, one band crosses $E_F$ along $\Gamma$-K-M at $k_z\sim$ 0 ($\alpha$ band), while two bands are very close to $E_F$ at $k_z\sim$ 0.25 $b^*$, including one shallow electron band at $\overline{\Gamma}$ poking through $E_F$ ($\beta$ band), and another hole-like band near $k_x\sim$ 0.4 {\AA}$^{-1}$ ($\gamma$ band). The calculated energy of band $\gamma$ is slightly lower than the experimental value, likely due to inaccuracies in DFT calculations. Although the probed $k$ space for one photon energy is expected to be a curved $k_x-k_z$ line for ideal free-electron final states (Fig.~\ref{Fig1}(d)), we used a fixed $k_z$ for quantitative comparison with experiment. Such simplified treatment is often employed in practice; further comparisons considering these curved $k_x-k_z$ lines are shown in Fig. S3 in \cite{supplementary}.

\begin{figure}[ht]
\centering
\includegraphics[width=1.\columnwidth]{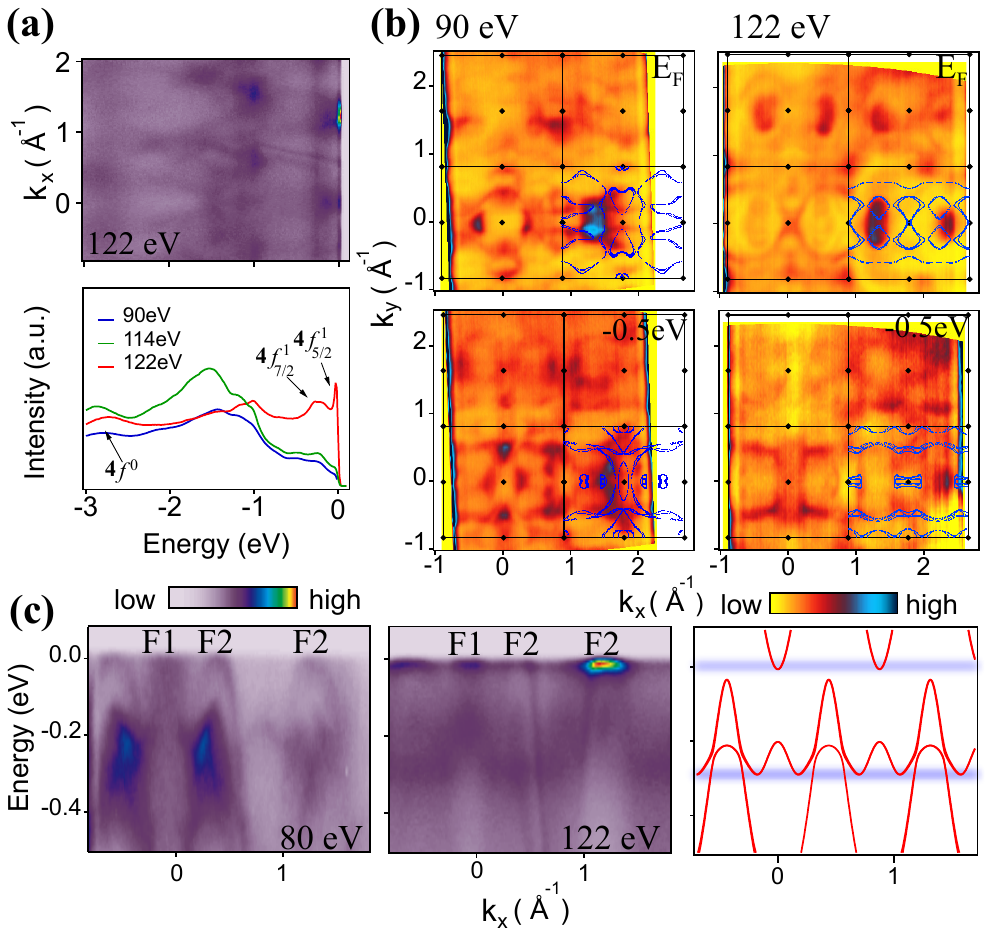}
\caption{(Color online). (a) Resonant ARPES spectrum (top) and the integrated energy distribution curves (EDCs) in comparison with off-resonant data (bottom). (b) Constant energy maps at $E_F$ (top panels) and -0.5 eV (bottom panels) at 90 eV (left, $k_z \sim$ 0) and 122 eV (right, $k_z \sim$ 0.25 $b^*$). The calculations are shown in the bottom right BZ (blue dots). (c) Comparison of the 80 eV and 122 eV spectra along $k_x$ at the same $k_z$ and $k_y$ ($k_z \sim$ 0.25 $b^*$, $k_y$ = 0). The rightmost panel shows the conduction bands from the localized $4f$ calculation (red curves) and the expected $4f^1$ peaks (diffuse blue lines): their hybridization leads to the dispersive $4f$ bands observed experimentally.
}
\label{Fig2}
\end{figure}

To probe the Ce $4f$ electrons, we utilized resonant photoemission at 122 eV. While the off-resonant spectra are dominated by non-$4f$ conduction electrons, the $4f$ spectral weight is substantially enhanced at the resonance condition \cite{sekiyama2000probing}. The resonant scan (Fig.~\ref{Fig2}(a)) reveals a broad peak at -2.7 eV (localized $4f^0$), and sharp $4f^1_{5/2}$ and $4f^1_{7/2}$ peaks at $E_F$ and -0.3 eV, respectively. The sharpness and large intensity of the $4f^1$ peaks with respect to $4f^0$ suggests that the Kondo effect is active at this temperature (17 K) \cite{jw2005kondo,Li2021Charge}. Constant energy maps at $E_F$ and -0.5 eV from two representative photon energies are summarized in Fig.~\ref{Fig2}(b), together with localized $4f$ calculations. At $E$ = -0.5 eV, both the $k_x-k_y$ maps at 90 eV and 122 eV feature flat segments running along $k_x$ (perpendicular to the Ce chains), and small pockets near the BZ boundaries, in reasonable agreement with calculations. The FS maps exhibit similarly wiggling bands extending along $k_x$, but the patterns deviate considerably from the localized $4f$ calculations. The difference between -0.5 eV and $E_F$ can be explained by the simple hybridized band picture based on the periodic Anderson model (PAM) \cite{denlinger2001comparative}: the $4f$ electrons contribute to the quasiparticle band dispersion only in the vicinity of $E_F$, via emergent Kondo peaks and their hybridization with conduction bands. This can be better illustrated in Fig.~\ref{Fig2}(c), where we compare the spectra near $E_F$ taken with 80 and 122 eV (on resonance) photons at the same $k_z$ and $k_y$, with the localized $4f$ calculation. Since simple 'itinerant $4f$'  calculations using DFT are unapt to explain the experimental results due to the strong local correlations from Ce $4f$ electrons (Fig. S2 in \cite{supplementary}), we adopt the aforementioned hybridized band picture to interpret our data. Here the dispersive $4f$ bands near $E_F$ result from the periodic arrangement of $4f$ sites which turns the local Kondo singlets (diffuse blue lines in Fig.~\ref{Fig2}(c)) into slowly propagating Bloch states, via hybridization with conduction bands. Experimentally, two types of symmetry-inequivalent $4f$ bands can be identified at $E_F$ from the resonance enhancement, one at $k_x\sim$ 0 (F1) corresponding to crossing with the electron band $\beta$ (Fig.~\ref{Fig1}(f)), the other at $k_x \sim \pm$0.4 {\AA}$^{-1}$ and 1.2 {\AA}$^{-1}$ (F2), due to crossing with the hole band $\gamma$. The $c-f$ hybridization can be best seen from F2 near 1.2 Å$^{-1}$, where the corresponding conduction band exhibits bending near $E_F$ (80 eV data) and the intense F2 peak suddenly loses most of its weight at the crossing point with the conduction band (122 eV data), characteristic of the $c-f$ hybridization. We note that the F1 and F2 bands in Fig.~\ref{Fig2}(c) are not perfectly periodic in experiments, likely due to the curved $k_x-k_z$ line probed by ARPES (Fig.~\ref{Fig1}(d)) and different photoemission matrix elements.

\begin{figure}[ht]
\includegraphics[width=1.\columnwidth]{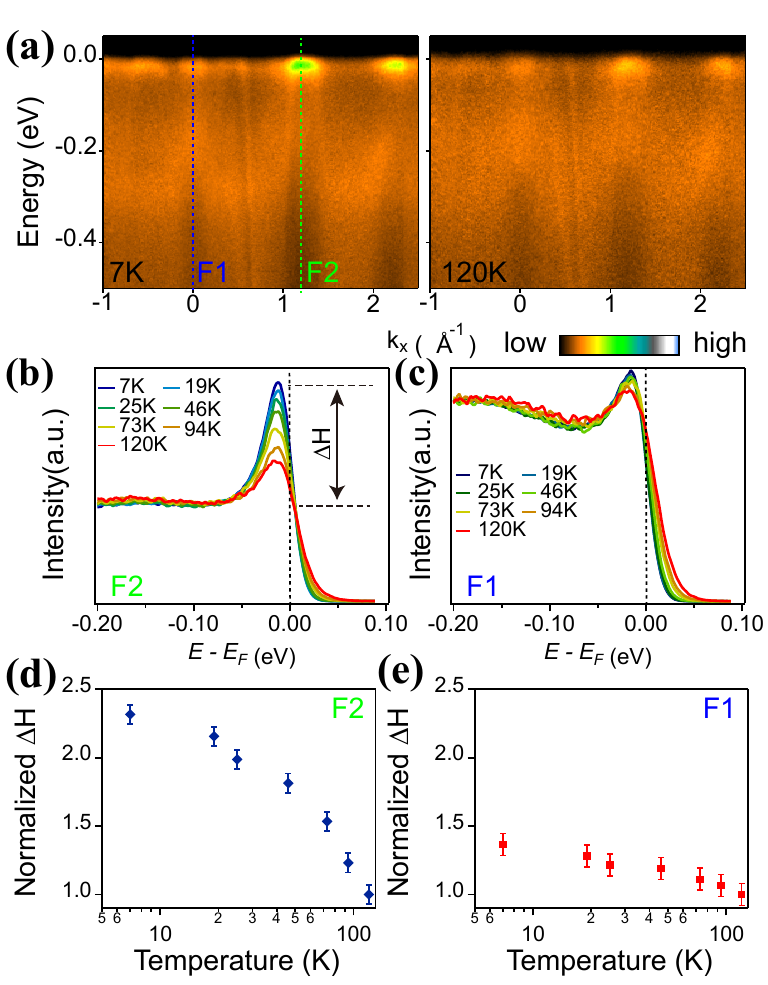}
\caption{(Color online). (a) Resonant ARPES spectra at 7 K and 120 K ($k_y$ = 0). (b,c) Temperature evolution of EDCs corresponding to green (F2 peak, (b)) and blue (F1 peak, (c)) dashed lines in (a). (d,e) Background-subtracted peak height $\Delta$H (defined in (b)) as a function of temperature for F2 (d) and F1 (e), normalized to the values at 120 K.
}
\label{Fig3}
\end{figure}

Fig.~\ref{Fig3}(a) shows the temperature-dependent resonant ARPES spectra. While there is obvious intensity reduction with increasing temperature for the F2 peak (Fig.~\ref{Fig3}(b)), the F1 peak changes much less with temperature (Fig.~\ref{Fig3}(c)). For quantitative analysis, we plot the background-subtracted peak height as a function of temperature in Fig.~\ref{Fig3}(d,e) (see also Fig. S5 in \cite{supplementary}). The F2 intensity roughly follows the $-\log(T)$ behavior (expected for a Kondo system) in the measurement temperature range. The peak is observable up to high temperature (intensity reduction $\sim60\%$ at 120 K compared to 7 K), much higher than the Kondo temperature $T_K \sim$ 20 K \cite{matsuoka2015ferromagnetic} and the transport coherence temperature $T^{*} \sim$ 80 K (Fig. S1 in \cite{supplementary}). This is consistent with the onset of $c-f$ hybridization at high temperature obtained from a recent ultrafast optical pump-probe measurement \cite{Qi2021Ultrafast}. This behavior is also similar to other HF compounds, e.g., CeCoIn$_5$ \cite{chen2017direct,jang2017evolution}, and it could be attributed to Kondo screening involving CEF excitations \cite{chen2017direct,jang2017evolution,kroha2003structure,kummer2015temperature}. Analysis of the magnetic susceptibility and inelastic neutron scattering (INS) suggests that the first excited CEF doublet in CeRh$_6$Ge$_4$ lies at $\sim$6 meV above the ground-state doublet and it hybridizes strongly with the conduction bands \cite{micheal2020}. This is also supported by a low-temperature Kadowaki-Woods ratio corresponding to a $4f$ ground state degeneracy $N$ = 4 \cite{shen2020strange}. Therefore, the F2 peak could contain contributions from the low-lying excited doublet, effectively enhancing its coherence temperature. In contrast, the decrease in the F1 intensity with increasing temperature is much less dramatic. Interestingly, the integrated peak intensity of F1 after background subtraction actually increases with temperature (Fig. S5 in \cite{supplementary}). Such an intensity increase could be caused by crossing with an electron-type conduction band (Fig.~\ref{Fig2}(c)), whose band bottom lies very close to the $4f$ band. This uncommon band crossing and hybridization leads to additional hybridized state(s) slightly above $E_F$, contributing to the integrated intensity at elevated temperatures (Fig. S6 in \cite{supplementary}). We note that a weak momentum-independent $4f$ band can also be observed near $E_F$ at low temperature (Fig.~\ref{Fig3}(a)), but it is almost absent at 120 K, exhibiting a different temperature dependence from the F1 and F2 peaks (Fig. S7 in \cite{supplementary}).

\begin{figure}[ht]
\includegraphics[width=1.\columnwidth]{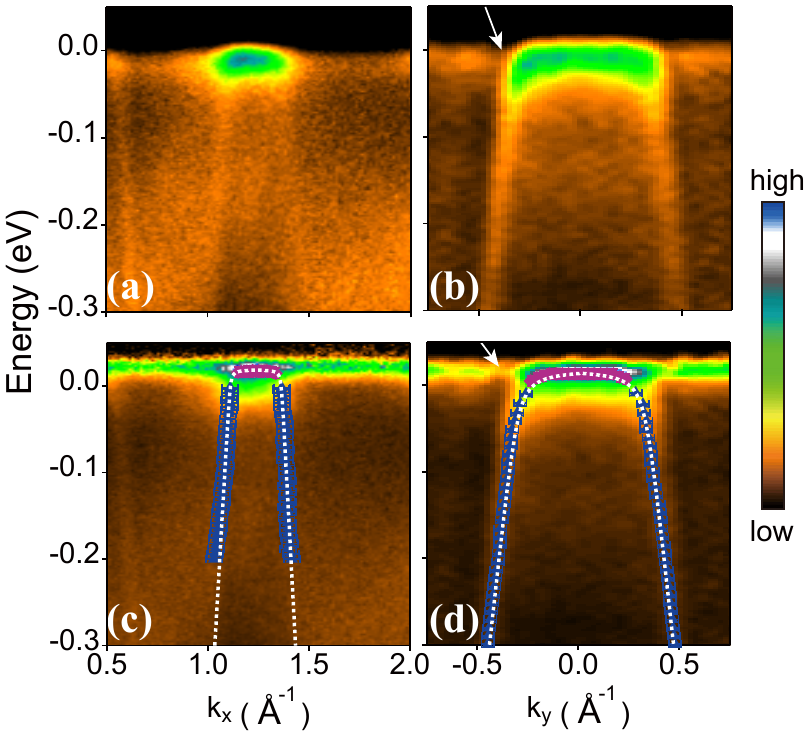}
\caption {(Color online). (a,b) Resonant ARPES spectra across the F2 peak along directions perpendicular ($k_x$, a) and parallel ($k_y$, b) to the Ce chains. (c,d) ARPES spectra in (a,b) divided by RC-FDD. Purple diamonds (blue triangles) with error bars indicate extracted peak positions from EDC (MDC) analysis. White dotted lines are model fittings. The arrows in (b,d) highlight the hybridization gap (= 2$V$) along $k_y$.
}
\label{Fig4}
\end{figure}

As the F2 bands make large contributions to the FS (Fig.~\ref{Fig2}), we performed a detailed analysis of its band dispersion along directions both perpendicular and parallel to the Ce chains (Fig.~\ref{Fig4}(a,b)). The results reveals a clear difference in the magnitude of the hybridization-induced band bending along two directions (Fig. S8 in \cite{supplementary}), implying different $c-f$ hybridization strengths. To recover the full spectral function near $E_F$, we divided the ARPES spectra by the resolution-convoluted Fermi-Dirac distribution (RC-FDD) \cite{chen2017direct,ehm2007high,im2008direct}, as shown in Fig.~\ref{Fig4}(c,d). While the recovered $4f$ band is quite flat along $k_x$ (perpendicular to the chain) without a clear signature of the $c-f$ hybridization gap, the quasiparticle dispersion along $k_y$ (parallel to the chain) shows a more pronounced bending of the reversed-U shaped band, as well as a clear hybridization gap (arrows in Fig.~\ref{Fig4}(b,d)). This implies that the $c-f$ hybridization could be much stronger along $k_y$ compared to $k_x$. To estimate the hybridization strength, we adopted the hybridized band approach discussed above, where the band dispersion is described by
\begin{equation}\label{1}
  E^\pm(k)=\frac{\overline{\varepsilon}_f+\varepsilon_k\pm\sqrt{(\overline{\varepsilon}_f-\varepsilon_k)^2+4V^2}}{2}.
\end{equation}
Here $\overline{\varepsilon}_f$ and $\varepsilon_k$ are the energies of the renormalized $4f$ peak and conduction band respectively, and $V$ is the strength of the $c-f$ hybridization. For simplicity, we used a linear dispersion to simulate $\varepsilon_k$ near $E_F$ and constrained our analysis to the lower branch $E^-(k)$. The experimental dispersion can be extracted from simultaneous analysis of the momentum-distribution curves (MDCs) and EDCs (Fig.~\ref{Fig4}(c,d)). The experimental dispersion can be reasonably described by this model with fitted values of $V$ $\approx$ 62 meV and 20 meV along $k_y$ and $k_x$, respectively (Fig. S9 in \cite{supplementary}). The uncertainty of $V$ is $\sim \pm10$ meV along $k_y$ and slightly larger along $k_x$ due to weaker bands with smaller bending. Another source of uncertainty in the estimation of $V$ is related to the RC-FDD division: since the ARPES energy resolution ($\sim20$ meV) is larger than 4$k_B$$T$, the recovered $4f$ bands above $E_F$ could be pushed slightly above the real positions \cite{jang2017evolution}, resulting in possible inaccuracy in $V$. However, since $V$ along $k_y$ is much larger than the possible energy shift, this complication should not affect our main conclusion: $V$ is obviously larger along $k_y$ compared to along $k_x$. Such an anisotropic $V$ is also manifested in the raw data via the presence (absence) of a clear $c-f$ hybridization gap along $k_y$ ($k_x$) (Fig.~\ref{Fig4}(a,b)).

The momentum anisotropy in $V$ can be attributed to the crystal structure: the $c-f$ hybridization should be dominated by electronic couplings between Ce and twelve nearest neighbor Rh and Ge atoms, which form small hexagons in the $a-b$ plane halfway between Ce atoms along the chains. Since the charge distribution of the $4f$ CEF ground state (most likely $\vert\pm1/2\rangle$) is mainly along the $c$ axis \cite{micheal2020}, addition of these hybridization channels could lead to stronger $c-f$ hybridization along the chain. As the Ruderman-Kittel-Kasuya-Yosida (RKKY) exchange interaction is realized by the same local electronic couplings between Ce and Rh/Ge atoms \cite{HF2007coleman}, our observed anisotropy in $V$ should be directly connected to the dominant RKKY interaction along the chain proposed in this system \cite{shen2020strange}. While such quasi-1D magnetic anisotropy remains to be confirmed experimentally, theoretical studies indicate that such 1D magnetism can suppress the first-order transition that normally occurs in an isotropic ferromagnet \cite{shen2020strange,Komijani2018Model}. Avoiding such a first-order transition is essential for the observation of the FM QCP \cite{brando2016metallic}. In another recent theoretical paper \cite{kirkpatrick2020ferromagnetic}, the strong spin-orbit coupling (SOC) in the non-centrosymmetric structure was proposed as an alternative mechanism for the observed FM QCP. While SOC is clearly strong for the $4f$ bands, as evidenced by the SOC-split $4f^{1}_{5/2}$ and $4f^{1}_{7/2}$ peaks, we found that the SOC splitting of non-$4f$ conduction bands (from Rh and Ge) is too small to resolve in our experiment. Future theoretical studies are needed to quantitatively understand the anisotropic $c-f$ hybridization and the possible role of SOC.

Our results therefore provide spectroscopic evidence for a strong Kondo effect and a pronounced $k$-space anisotropy in the $4f$ spectral weight and $c-f$ hybridization strength $V$ well above $T_C$ in the FM quantum critical metal CeRh$_6$Ge$_4$. The electronic structure is three-dimensional despite the chain-like arrangement of Ce, but the resulting $k$-space anisotropy in $V$ can be naturally linked to the proposed quasi-1D magnetic anisotropy, which could be key for the observed FM QCP \cite{shen2020strange}. Since the electronic structure of FM Kondo-lattice systems, particularly their momentum-resolved $4f$ bands, are much less studied by ARPES \cite{Yano2007Three,generalov2017insight,yamaoka2017electronic}, compared to their AFM counterparts, our results can be useful for a basic understanding of these materials. It is interesting to note that FM rare-earth Kondo-lattice systems are rare compared to AFM systems \cite{hafner2019kondo,Ahamed2018PRB}. They also show Kondo temperatures typically close to (or slightly higher than) $T_C$ \cite{hafner2019kondo}, and the ordered moment in the FM phase is usually small. Theoretical studies have indicated possible coexistence of Kondo screening and FM ordering over a large phase space \cite{Li2010PRB,Bernhard2015PRB}.

It would be interesting in future to track the temperature evolution of the $4f$ bands across $T_C$, and to examine the possible appearance of a 'small FS' (excluding $4f$ electrons) deep inside the FM phase \cite{shen2020strange,smidman2018interplay}. Recent measurements from quantum oscillations (QOs) showed that the observed FS in the FM phase of CeRh$_6$Ge$_4$ is close to the localized $4f$ calculation \cite{An2020QO}, although some discrepancy is still present implying possibly coexisting (dynamic) Kondo effect \cite{hu2020quantum}. It is intriguing that the AFM Kondo-lattice system CeRhIn$_5$ (with the local quantum criticality involving a simultaneous Kondo breakdown) shares interesting similarities with CeRh$_6$Ge$_4$: QO measurements revealed a 'small FS' well below the N\'eel temperature $T_N$ \cite{shishido2002JPSJ,Harrison2004PRL}, while ARPES measurement detected hybridized $4f$ bands well above $T_N$ \cite{chen2018band}. Since the translation symmetry is preserved across $T_C$ in FM systems (unlike in AFM systems), such a temperature-dependent study in CeRh$_6$Ge$_4$ may shed light on the delicate interplay between the (dynamic) Kondo effect and FM order near a QCP \cite{hu2020quantum}.

\par This work is supported by the National Key R$\&$D Program of China (Grant No. 2017YFA0303100, 2016YFA0300203), the National Science Foundation of China (No. 11674280, 12034017), the Key R$\&$D Program of Zhejiang Province, China (2021C01002) and the Fundamental Research Funds for the Central Universities. The ALS is supported by the Office of Basic Energy Sciences of the U.S. DOE under Contract No. DE-AC02-05CH11231.


\begin{thebibliography}{44}%
\makeatletter
\providecommand \@ifxundefined [1]{%
 \@ifx{#1\undefined}
}%
\providecommand \@ifnum [1]{%
 \ifnum #1\expandafter \@firstoftwo
 \else \expandafter \@secondoftwo
 \fi
}%
\providecommand \@ifx [1]{%
 \ifx #1\expandafter \@firstoftwo
 \else \expandafter \@secondoftwo
 \fi
}%
\providecommand \natexlab [1]{#1}%
\providecommand \enquote  [1]{``#1''}%
\providecommand \bibnamefont  [1]{#1}%
\providecommand \bibfnamefont [1]{#1}%
\providecommand \citenamefont [1]{#1}%
\providecommand \href@noop [0]{\@secondoftwo}%
\providecommand \href [0]{\begingroup \@sanitize@url \@href}%
\providecommand \@href[1]{\@@startlink{#1}\@@href}%
\providecommand \@@href[1]{\endgroup#1\@@endlink}%
\providecommand \@sanitize@url [0]{\catcode `\\12\catcode `\$12\catcode
  `\&12\catcode `\#12\catcode `\^12\catcode `\_12\catcode `\%12\relax}%
\providecommand \@@startlink[1]{}%
\providecommand \@@endlink[0]{}%
\providecommand \url  [0]{\begingroup\@sanitize@url \@url }%
\providecommand \@url [1]{\endgroup\@href {#1}{\urlprefix }}%
\providecommand \urlprefix  [0]{URL }%
\providecommand \Eprint [0]{\href }%
\providecommand \doibase [0]{https://doi.org/}%
\providecommand \selectlanguage [0]{\@gobble}%
\providecommand \bibinfo  [0]{\@secondoftwo}%
\providecommand \bibfield  [0]{\@secondoftwo}%
\providecommand \translation [1]{[#1]}%
\providecommand \BibitemOpen [0]{}%
\providecommand \bibitemStop [0]{}%
\providecommand \bibitemNoStop [0]{.\EOS\space}%
\providecommand \EOS [0]{\spacefactor3000\relax}%
\providecommand \BibitemShut  [1]{\csname bibitem#1\endcsname}%
\let\auto@bib@innerbib\@empty
\bibitem [{\citenamefont {Shen}\ \emph {et~al.}(2020)\citenamefont {Shen},
  \citenamefont {Zhang}, \citenamefont {Komijani}, \citenamefont {Nicklas},
  \citenamefont {Borth}, \citenamefont {Wang}, \citenamefont {Chen},
  \citenamefont {Nie}, \citenamefont {Li}, \citenamefont {Lu}, \citenamefont
  {Lee}, \citenamefont {Smidman}, \citenamefont {Steglich}, \citenamefont
  {Coleman},\ and\ \citenamefont {Yuan}}]{shen2020strange}%
  \BibitemOpen
  \bibfield  {author} {\bibinfo {author} {\bibfnamefont {B.}~\bibnamefont
  {Shen}}, \bibinfo {author} {\bibfnamefont {Y.}~\bibnamefont {Zhang}},
  \bibinfo {author} {\bibfnamefont {Y.}~\bibnamefont {Komijani}}, \bibinfo
  {author} {\bibfnamefont {M.}~\bibnamefont {Nicklas}}, \bibinfo {author}
  {\bibfnamefont {R.}~\bibnamefont {Borth}}, \bibinfo {author} {\bibfnamefont
  {A.}~\bibnamefont {Wang}}, \bibinfo {author} {\bibfnamefont {Y.}~\bibnamefont
  {Chen}}, \bibinfo {author} {\bibfnamefont {Z.}~\bibnamefont {Nie}}, \bibinfo
  {author} {\bibfnamefont {R.}~\bibnamefont {Li}}, \bibinfo {author}
  {\bibfnamefont {X.}~\bibnamefont {Lu}}, \bibinfo {author} {\bibfnamefont
  {H.}~\bibnamefont {Lee}}, \bibinfo {author} {\bibfnamefont {M.}~\bibnamefont
  {Smidman}}, \bibinfo {author} {\bibfnamefont {F.}~\bibnamefont {Steglich}},
  \bibinfo {author} {\bibfnamefont {P.}~\bibnamefont {Coleman}},\ and\ \bibinfo
  {author} {\bibfnamefont {H.}~\bibnamefont {Yuan}},\ }\href
  {https://doi.org/10.1038/s41586-020-2052-z} {\bibfield  {journal} {\bibinfo
  {journal} {Nature}\ }\textbf {\bibinfo {volume} {579}},\ \bibinfo {pages}
  {51} (\bibinfo {year} {2020})}\BibitemShut {NoStop}%
\bibitem [{\citenamefont {Gegenwart}\ \emph {et~al.}(2008)\citenamefont
  {Gegenwart}, \citenamefont {Si},\ and\ \citenamefont
  {Steglich}}]{gegenwart2008quantum}%
  \BibitemOpen
  \bibfield  {author} {\bibinfo {author} {\bibfnamefont {P.}~\bibnamefont
  {Gegenwart}}, \bibinfo {author} {\bibfnamefont {Q.}~\bibnamefont {Si}},\ and\
  \bibinfo {author} {\bibfnamefont {F.}~\bibnamefont {Steglich}},\ }\href
  {https://doi.org/10.1038/nphys892} {\bibfield  {journal} {\bibinfo  {journal}
  {Nature physics}\ }\textbf {\bibinfo {volume} {4}},\ \bibinfo {pages} {186}
  (\bibinfo {year} {2008})}\BibitemShut {NoStop}%
\bibitem [{\citenamefont {Belitz}\ \emph {et~al.}(1999)\citenamefont {Belitz},
  \citenamefont {Kirkpatrick},\ and\ \citenamefont {Vojta}}]{Belitz1999first}%
  \BibitemOpen
  \bibfield  {author} {\bibinfo {author} {\bibfnamefont {D.}~\bibnamefont
  {Belitz}}, \bibinfo {author} {\bibfnamefont {T.~R.}\ \bibnamefont
  {Kirkpatrick}},\ and\ \bibinfo {author} {\bibfnamefont {T.}~\bibnamefont
  {Vojta}},\ }\href {https://doi.org/10.1103/PhysRevLett.82.4707} {\bibfield
  {journal} {\bibinfo  {journal} {Phys. Rev. Lett.}\ }\textbf {\bibinfo
  {volume} {82}},\ \bibinfo {pages} {4707} (\bibinfo {year}
  {1999})}\BibitemShut {NoStop}%
\bibitem [{\citenamefont {Brando}\ \emph {et~al.}(2016)\citenamefont {Brando},
  \citenamefont {Belitz}, \citenamefont {Grosche},\ and\ \citenamefont
  {Kirkpatrick}}]{brando2016metallic}%
  \BibitemOpen
  \bibfield  {author} {\bibinfo {author} {\bibfnamefont {M.}~\bibnamefont
  {Brando}}, \bibinfo {author} {\bibfnamefont {D.}~\bibnamefont {Belitz}},
  \bibinfo {author} {\bibfnamefont {F.~M.}\ \bibnamefont {Grosche}},\ and\
  \bibinfo {author} {\bibfnamefont {T.~R.}\ \bibnamefont {Kirkpatrick}},\
  }\href {https://doi.org/10.1103/RevModPhys.88.025006} {\bibfield  {journal}
  {\bibinfo  {journal} {Rev. Mod. Phys.}\ }\textbf {\bibinfo {volume} {88}},\
  \bibinfo {pages} {025006} (\bibinfo {year} {2016})}\BibitemShut {NoStop}%
\bibitem [{\citenamefont {Chubukov}\ \emph {et~al.}(2004)\citenamefont
  {Chubukov}, \citenamefont {P\'epin},\ and\ \citenamefont
  {Rech}}]{Chubukov2004instability}%
  \BibitemOpen
  \bibfield  {author} {\bibinfo {author} {\bibfnamefont {A.~V.}\ \bibnamefont
  {Chubukov}}, \bibinfo {author} {\bibfnamefont {C.}~\bibnamefont {P\'epin}},\
  and\ \bibinfo {author} {\bibfnamefont {J.}~\bibnamefont {Rech}},\ }\href
  {https://doi.org/10.1103/PhysRevLett.92.147003} {\bibfield  {journal}
  {\bibinfo  {journal} {Phys. Rev. Lett.}\ }\textbf {\bibinfo {volume} {92}},\
  \bibinfo {pages} {147003} (\bibinfo {year} {2004})}\BibitemShut {NoStop}%
\bibitem [{\citenamefont {Stewart}(2001)}]{RevModPhys.73.797}%
  \BibitemOpen
  \bibfield  {author} {\bibinfo {author} {\bibfnamefont {G.~R.}\ \bibnamefont
  {Stewart}},\ }\href {https://doi.org/10.1103/RevModPhys.73.797} {\bibfield
  {journal} {\bibinfo  {journal} {Rev. Mod. Phys.}\ }\textbf {\bibinfo {volume}
  {73}},\ \bibinfo {pages} {797} (\bibinfo {year} {2001})}\BibitemShut
  {NoStop}%
\bibitem [{\citenamefont {L\"ohneysen}\ \emph {et~al.}(2007)\citenamefont
  {L\"ohneysen}, \citenamefont {Rosch}, \citenamefont {Vojta},\ and\
  \citenamefont {W\"olfle}}]{Lohneysen2007Fermi}%
  \BibitemOpen
  \bibfield  {author} {\bibinfo {author} {\bibfnamefont {H.~v.}\ \bibnamefont
  {L\"ohneysen}}, \bibinfo {author} {\bibfnamefont {A.}~\bibnamefont {Rosch}},
  \bibinfo {author} {\bibfnamefont {M.}~\bibnamefont {Vojta}},\ and\ \bibinfo
  {author} {\bibfnamefont {P.}~\bibnamefont {W\"olfle}},\ }\href
  {https://doi.org/10.1103/RevModPhys.79.1015} {\bibfield  {journal} {\bibinfo
  {journal} {Rev. Mod. Phys.}\ }\textbf {\bibinfo {volume} {79}},\ \bibinfo
  {pages} {1015} (\bibinfo {year} {2007})}\BibitemShut {NoStop}%
\bibitem [{\citenamefont {Steppke}\ \emph {et~al.}(2013)\citenamefont
  {Steppke}, \citenamefont {K\"uchler}, \citenamefont {Lausberg}, \citenamefont
  {Lengyel}, \citenamefont {Steinke}, \citenamefont {Borth}, \citenamefont
  {L\"uhmann}, \citenamefont {Krellner}, \citenamefont {Nicklas}, \citenamefont
  {Geibel}, \citenamefont {Steglich},\ and\ \citenamefont
  {Brando}}]{steppke2013ferromagnetic}%
  \BibitemOpen
  \bibfield  {author} {\bibinfo {author} {\bibfnamefont {A.}~\bibnamefont
  {Steppke}}, \bibinfo {author} {\bibfnamefont {R.}~\bibnamefont {K\"uchler}},
  \bibinfo {author} {\bibfnamefont {S.}~\bibnamefont {Lausberg}}, \bibinfo
  {author} {\bibfnamefont {E.}~\bibnamefont {Lengyel}}, \bibinfo {author}
  {\bibfnamefont {L.}~\bibnamefont {Steinke}}, \bibinfo {author} {\bibfnamefont
  {R.}~\bibnamefont {Borth}}, \bibinfo {author} {\bibfnamefont
  {T.}~\bibnamefont {L\"uhmann}}, \bibinfo {author} {\bibfnamefont
  {C.}~\bibnamefont {Krellner}}, \bibinfo {author} {\bibfnamefont
  {M.}~\bibnamefont {Nicklas}}, \bibinfo {author} {\bibfnamefont
  {C.}~\bibnamefont {Geibel}}, \bibinfo {author} {\bibfnamefont
  {F.}~\bibnamefont {Steglich}},\ and\ \bibinfo {author} {\bibfnamefont
  {M.}~\bibnamefont {Brando}},\ }\href
  {https://doi.org/10.1126/science.1230583} {\bibfield  {journal} {\bibinfo
  {journal} {Science}\ }\textbf {\bibinfo {volume} {339}},\ \bibinfo {pages}
  {933} (\bibinfo {year} {2013})}\BibitemShut {NoStop}%
\bibitem [{\citenamefont {Krellner}\ \emph {et~al.}(2011)\citenamefont
  {Krellner}, \citenamefont {Lausberg}, \citenamefont {Steppke}, \citenamefont
  {Brando}, \citenamefont {Pedrero}, \citenamefont {Pfau}, \citenamefont
  {Tenc\'e}, \citenamefont {Rosner}, \citenamefont {Steglich},\ and\
  \citenamefont {Geibel}}]{krellner2011ferromagnetic}%
  \BibitemOpen
  \bibfield  {author} {\bibinfo {author} {\bibfnamefont {C.}~\bibnamefont
  {Krellner}}, \bibinfo {author} {\bibfnamefont {S.}~\bibnamefont {Lausberg}},
  \bibinfo {author} {\bibfnamefont {A.}~\bibnamefont {Steppke}}, \bibinfo
  {author} {\bibfnamefont {M.}~\bibnamefont {Brando}}, \bibinfo {author}
  {\bibfnamefont {L.}~\bibnamefont {Pedrero}}, \bibinfo {author} {\bibfnamefont
  {H.}~\bibnamefont {Pfau}}, \bibinfo {author} {\bibfnamefont {S.}~\bibnamefont
  {Tenc\'e}}, \bibinfo {author} {\bibfnamefont {H.}~\bibnamefont {Rosner}},
  \bibinfo {author} {\bibfnamefont {F.}~\bibnamefont {Steglich}},\ and\
  \bibinfo {author} {\bibfnamefont {C.}~\bibnamefont {Geibel}},\ }\href
  {https://doi.org/10.1088/1367-2630/13/10/103014} {\bibfield  {journal}
  {\bibinfo  {journal} {New Journal of Physics}\ }\textbf {\bibinfo {volume}
  {13}},\ \bibinfo {pages} {103014} (\bibinfo {year} {2011})}\BibitemShut
  {NoStop}%
\bibitem [{\citenamefont {Wang}\ \emph {et~al.}(2017)\citenamefont {Wang},
  \citenamefont {Fu}, \citenamefont {Sun}, \citenamefont {Liu}, \citenamefont
  {Yi}, \citenamefont {Yi}, \citenamefont {Luo}, \citenamefont {Dai},
  \citenamefont {Liu}, \citenamefont {Matsushita}, \citenamefont {Yamaura},
  \citenamefont {Lu}, \citenamefont {Cheng}, \citenamefont {Yang},
  \citenamefont {Shi},\ and\ \citenamefont {Luo}}]{wang2017heavy}%
  \BibitemOpen
  \bibfield  {author} {\bibinfo {author} {\bibfnamefont {L.}~\bibnamefont
  {Wang}}, \bibinfo {author} {\bibfnamefont {Z.}~\bibnamefont {Fu}}, \bibinfo
  {author} {\bibfnamefont {J.}~\bibnamefont {Sun}}, \bibinfo {author}
  {\bibfnamefont {M.}~\bibnamefont {Liu}}, \bibinfo {author} {\bibfnamefont
  {W.}~\bibnamefont {Yi}}, \bibinfo {author} {\bibfnamefont {C.}~\bibnamefont
  {Yi}}, \bibinfo {author} {\bibfnamefont {Y.}~\bibnamefont {Luo}}, \bibinfo
  {author} {\bibfnamefont {Y.}~\bibnamefont {Dai}}, \bibinfo {author}
  {\bibfnamefont {G.}~\bibnamefont {Liu}}, \bibinfo {author} {\bibfnamefont
  {Y.}~\bibnamefont {Matsushita}}, \bibinfo {author} {\bibfnamefont
  {K.}~\bibnamefont {Yamaura}}, \bibinfo {author} {\bibfnamefont
  {L.}~\bibnamefont {Lu}}, \bibinfo {author} {\bibfnamefont {J.-G.}\
  \bibnamefont {Cheng}}, \bibinfo {author} {\bibfnamefont {Y.-F.}\ \bibnamefont
  {Yang}}, \bibinfo {author} {\bibfnamefont {Y.}~\bibnamefont {Shi}},\ and\
  \bibinfo {author} {\bibfnamefont {J.}~\bibnamefont {Luo}},\ }\href
  {https://doi.org/10.1038/s41535-017-0040-9} {\bibfield  {journal} {\bibinfo
  {journal} {npj Quantum Materials}\ }\textbf {\bibinfo {volume} {2}},\
  \bibinfo {pages} {1} (\bibinfo {year} {2017})}\BibitemShut {NoStop}%
\bibitem [{\citenamefont {Cheng}\ \emph {et~al.}(2019)\citenamefont {Cheng},
  \citenamefont {Wang}, \citenamefont {Xu}, \citenamefont {Yang}, \citenamefont
  {Zhu}, \citenamefont {Ke}, \citenamefont {Lu}, \citenamefont {Xia},
  \citenamefont {Wang}, \citenamefont {Shi}, \citenamefont {Yang},\ and\
  \citenamefont {Luo}}]{Cheng2019PRM}%
  \BibitemOpen
  \bibfield  {author} {\bibinfo {author} {\bibfnamefont {K.}~\bibnamefont
  {Cheng}}, \bibinfo {author} {\bibfnamefont {L.}~\bibnamefont {Wang}},
  \bibinfo {author} {\bibfnamefont {Y.}~\bibnamefont {Xu}}, \bibinfo {author}
  {\bibfnamefont {F.}~\bibnamefont {Yang}}, \bibinfo {author} {\bibfnamefont
  {H.}~\bibnamefont {Zhu}}, \bibinfo {author} {\bibfnamefont {J.}~\bibnamefont
  {Ke}}, \bibinfo {author} {\bibfnamefont {X.}~\bibnamefont {Lu}}, \bibinfo
  {author} {\bibfnamefont {Z.}~\bibnamefont {Xia}}, \bibinfo {author}
  {\bibfnamefont {J.}~\bibnamefont {Wang}}, \bibinfo {author} {\bibfnamefont
  {Y.}~\bibnamefont {Shi}}, \bibinfo {author} {\bibfnamefont {Y.}~\bibnamefont
  {Yang}},\ and\ \bibinfo {author} {\bibfnamefont {Y.}~\bibnamefont {Luo}},\
  }\href {https://doi.org/10.1103/PhysRevMaterials.3.021402} {\bibfield
  {journal} {\bibinfo  {journal} {Phys. Rev. Materials}\ }\textbf {\bibinfo
  {volume} {3}},\ \bibinfo {pages} {021402} (\bibinfo {year}
  {2019})}\BibitemShut {NoStop}%
\bibitem [{\citenamefont {Miao}\ \emph {et~al.}(2020)\citenamefont {Miao},
  \citenamefont {Liu}, \citenamefont {Xu}, \citenamefont {Kotta}, \citenamefont
  {Kang}, \citenamefont {Ran}, \citenamefont {Paglione}, \citenamefont
  {Kotliar}, \citenamefont {Butch}, \citenamefont {Denlinger},\ and\
  \citenamefont {Wray}}]{Miao2020PRL}%
  \BibitemOpen
  \bibfield  {author} {\bibinfo {author} {\bibfnamefont {L.}~\bibnamefont
  {Miao}}, \bibinfo {author} {\bibfnamefont {S.}~\bibnamefont {Liu}}, \bibinfo
  {author} {\bibfnamefont {Y.}~\bibnamefont {Xu}}, \bibinfo {author}
  {\bibfnamefont {E.~C.}\ \bibnamefont {Kotta}}, \bibinfo {author}
  {\bibfnamefont {C.-J.}\ \bibnamefont {Kang}}, \bibinfo {author}
  {\bibfnamefont {S.}~\bibnamefont {Ran}}, \bibinfo {author} {\bibfnamefont
  {J.}~\bibnamefont {Paglione}}, \bibinfo {author} {\bibfnamefont
  {G.}~\bibnamefont {Kotliar}}, \bibinfo {author} {\bibfnamefont {N.~P.}\
  \bibnamefont {Butch}}, \bibinfo {author} {\bibfnamefont {J.~D.}\ \bibnamefont
  {Denlinger}},\ and\ \bibinfo {author} {\bibfnamefont {L.~A.}\ \bibnamefont
  {Wray}},\ }\href {https://doi.org/10.1103/PhysRevLett.124.076401} {\bibfield
  {journal} {\bibinfo  {journal} {Phys. Rev. Lett.}\ }\textbf {\bibinfo
  {volume} {124}},\ \bibinfo {pages} {076401} (\bibinfo {year}
  {2020})}\BibitemShut {NoStop}%
\bibitem [{\citenamefont {Kotegawa}\ \emph {et~al.}(2019)\citenamefont
  {Kotegawa}, \citenamefont {Matsuoka}, \citenamefont {Uga}, \citenamefont
  {Takemura}, \citenamefont {Manago}, \citenamefont {Chikuchi}, \citenamefont
  {Sugawara}, \citenamefont {Tou},\ and\ \citenamefont
  {Harima}}]{kotegawa2019indication}%
  \BibitemOpen
  \bibfield  {author} {\bibinfo {author} {\bibfnamefont {H.}~\bibnamefont
  {Kotegawa}}, \bibinfo {author} {\bibfnamefont {E.}~\bibnamefont {Matsuoka}},
  \bibinfo {author} {\bibfnamefont {T.}~\bibnamefont {Uga}}, \bibinfo {author}
  {\bibfnamefont {M.}~\bibnamefont {Takemura}}, \bibinfo {author}
  {\bibfnamefont {M.}~\bibnamefont {Manago}}, \bibinfo {author} {\bibfnamefont
  {N.}~\bibnamefont {Chikuchi}}, \bibinfo {author} {\bibfnamefont
  {H.}~\bibnamefont {Sugawara}}, \bibinfo {author} {\bibfnamefont
  {H.}~\bibnamefont {Tou}},\ and\ \bibinfo {author} {\bibfnamefont
  {H.}~\bibnamefont {Harima}},\ }\href {https://doi.org/10.7566/JPSJ.88.093702}
  {\bibfield  {journal} {\bibinfo  {journal} {Journal of the Physical Society
  of Japan}\ }\textbf {\bibinfo {volume} {88}},\ \bibinfo {pages} {093702}
  (\bibinfo {year} {2019})}\BibitemShut {NoStop}%
\bibitem [{\citenamefont {Cao}\ and\ \citenamefont
  {Zhu}(2020)}]{cao2020pressure}%
  \BibitemOpen
  \bibfield  {author} {\bibinfo {author} {\bibfnamefont {C.}~\bibnamefont
  {Cao}}\ and\ \bibinfo {author} {\bibfnamefont {J.-X.}\ \bibnamefont {Zhu}},\
  }\href {https://arxiv.org/abs/2011.14256} {\bibfield  {journal} {\bibinfo
  {journal} {arXiv preprint arXiv:2011.14256}\ } (\bibinfo {year}
  {2020})}\BibitemShut {NoStop}%
\bibitem [{\citenamefont {Matsuoka}\ \emph {et~al.}(2015)\citenamefont
  {Matsuoka}, \citenamefont {Hondo}, \citenamefont {Fujii}, \citenamefont
  {Oshima}, \citenamefont {Sugawara}, \citenamefont {Sakurai}, \citenamefont
  {Ohta}, \citenamefont {Kneidinger}, \citenamefont {Salamakha}, \citenamefont
  {Michor},\ and\ \citenamefont {Bauer}}]{matsuoka2015ferromagnetic}%
  \BibitemOpen
  \bibfield  {author} {\bibinfo {author} {\bibfnamefont {E.}~\bibnamefont
  {Matsuoka}}, \bibinfo {author} {\bibfnamefont {C.}~\bibnamefont {Hondo}},
  \bibinfo {author} {\bibfnamefont {T.}~\bibnamefont {Fujii}}, \bibinfo
  {author} {\bibfnamefont {A.}~\bibnamefont {Oshima}}, \bibinfo {author}
  {\bibfnamefont {H.}~\bibnamefont {Sugawara}}, \bibinfo {author}
  {\bibfnamefont {T.}~\bibnamefont {Sakurai}}, \bibinfo {author} {\bibfnamefont
  {H.}~\bibnamefont {Ohta}}, \bibinfo {author} {\bibfnamefont {F.}~\bibnamefont
  {Kneidinger}}, \bibinfo {author} {\bibfnamefont {L.}~\bibnamefont
  {Salamakha}}, \bibinfo {author} {\bibfnamefont {H.}~\bibnamefont {Michor}},\
  and\ \bibinfo {author} {\bibfnamefont {E.}~\bibnamefont {Bauer}},\ }\href
  {https://doi.org/10.7566/JPSJ.84.073704} {\bibfield  {journal} {\bibinfo
  {journal} {Journal of the Physical Society of Japan}\ }\textbf {\bibinfo
  {volume} {84}},\ \bibinfo {pages} {073704} (\bibinfo {year}
  {2015})}\BibitemShut {NoStop}%
\bibitem [{\citenamefont {Hu}\ \emph {et~al.}(2020)\citenamefont {Hu},
  \citenamefont {Cai},\ and\ \citenamefont {Si}}]{hu2020quantum}%
  \BibitemOpen
  \bibfield  {author} {\bibinfo {author} {\bibfnamefont {H.}~\bibnamefont
  {Hu}}, \bibinfo {author} {\bibfnamefont {A.}~\bibnamefont {Cai}},\ and\
  \bibinfo {author} {\bibfnamefont {Q.}~\bibnamefont {Si}},\ }\href
  {https://arxiv.org/abs/2004.04679} {\bibfield  {journal} {\bibinfo  {journal}
  {arXiv preprint arXiv:2004.04679}\ } (\bibinfo {year} {2020})}\BibitemShut
  {NoStop}%
\bibitem [{sup()}]{supplementary}%
  \BibitemOpen
  \href@noop {} {}\bibinfo {note} {See online supplementary
  material}\BibitemShut {NoStop}%
\bibitem [{\citenamefont {Sekiyama}\ \emph {et~al.}(2000)\citenamefont
  {Sekiyama}, \citenamefont {Iwasaki}, \citenamefont {Matsuda}, \citenamefont
  {Saitoh}, \citenamefont {\^Onuki},\ and\ \citenamefont
  {Suga}}]{sekiyama2000probing}%
  \BibitemOpen
  \bibfield  {author} {\bibinfo {author} {\bibfnamefont {A.}~\bibnamefont
  {Sekiyama}}, \bibinfo {author} {\bibfnamefont {T.}~\bibnamefont {Iwasaki}},
  \bibinfo {author} {\bibfnamefont {K.}~\bibnamefont {Matsuda}}, \bibinfo
  {author} {\bibfnamefont {Y.}~\bibnamefont {Saitoh}}, \bibinfo {author}
  {\bibfnamefont {Y.}~\bibnamefont {\^Onuki}},\ and\ \bibinfo {author}
  {\bibfnamefont {S.}~\bibnamefont {Suga}},\ }\href
  {https://doi.org/10.1038/35000140} {\bibfield  {journal} {\bibinfo  {journal}
  {Nature}\ }\textbf {\bibinfo {volume} {403}},\ \bibinfo {pages} {396}
  (\bibinfo {year} {2000})}\BibitemShut {NoStop}%
\bibitem [{\citenamefont {Allen}(2005)}]{jw2005kondo}%
  \BibitemOpen
  \bibfield  {author} {\bibinfo {author} {\bibfnamefont {J.~W.}\ \bibnamefont
  {Allen}},\ }\href {https://doi.org/10.1143/JPSJ.74.34} {\bibfield  {journal}
  {\bibinfo  {journal} {Journal of the Physical Society of Japan}\ }\textbf
  {\bibinfo {volume} {74}},\ \bibinfo {pages} {34} (\bibinfo {year}
  {2005})}\BibitemShut {NoStop}%
\bibitem [{\citenamefont {Li}\ \emph {et~al.}(2021)\citenamefont {Li},
  \citenamefont {Lv}, \citenamefont {Fang}, \citenamefont {Guo}, \citenamefont
  {Wu}, \citenamefont {Wu}, \citenamefont {Shen}, \citenamefont {Nie},
  \citenamefont {Petaccia}, \citenamefont {Cao}, \citenamefont {Xu},\ and\
  \citenamefont {Liu}}]{Li2021Charge}%
  \BibitemOpen
  \bibfield  {author} {\bibinfo {author} {\bibfnamefont {P.}~\bibnamefont
  {Li}}, \bibinfo {author} {\bibfnamefont {B.}~\bibnamefont {Lv}}, \bibinfo
  {author} {\bibfnamefont {Y.}~\bibnamefont {Fang}}, \bibinfo {author}
  {\bibfnamefont {W.}~\bibnamefont {Guo}}, \bibinfo {author} {\bibfnamefont
  {Z.}~\bibnamefont {Wu}}, \bibinfo {author} {\bibfnamefont {Y.}~\bibnamefont
  {Wu}}, \bibinfo {author} {\bibfnamefont {D.}~\bibnamefont {Shen}}, \bibinfo
  {author} {\bibfnamefont {Y.}~\bibnamefont {Nie}}, \bibinfo {author}
  {\bibfnamefont {L.}~\bibnamefont {Petaccia}}, \bibinfo {author}
  {\bibfnamefont {C.}~\bibnamefont {Cao}}, \bibinfo {author} {\bibfnamefont
  {Z.~A.}\ \bibnamefont {Xu}},\ and\ \bibinfo {author} {\bibfnamefont
  {Y.}~\bibnamefont {Liu}},\ }\href {https://doi.org/10.1007/s11433-020-1642-2}
  {\bibfield  {journal} {\bibinfo  {journal} {Sci. China-Phys. Mech. Astron.}\
  }\textbf {\bibinfo {volume} {64}},\ \bibinfo {pages} {237412} (\bibinfo
  {year} {2021})}\BibitemShut {NoStop}%
\bibitem [{\citenamefont {Denlinger}\ \emph {et~al.}(2001)\citenamefont
  {Denlinger}, \citenamefont {Gweon}, \citenamefont {Allen}, \citenamefont
  {Olson}, \citenamefont {Maple}, \citenamefont {Sarrao}, \citenamefont
  {Armstrong}, \citenamefont {Fisk},\ and\ \citenamefont
  {Yamagami}}]{denlinger2001comparative}%
  \BibitemOpen
  \bibfield  {author} {\bibinfo {author} {\bibfnamefont {J.~D.}\ \bibnamefont
  {Denlinger}}, \bibinfo {author} {\bibfnamefont {G.-H.}\ \bibnamefont
  {Gweon}}, \bibinfo {author} {\bibfnamefont {J.~W.}\ \bibnamefont {Allen}},
  \bibinfo {author} {\bibfnamefont {C.~G.}\ \bibnamefont {Olson}}, \bibinfo
  {author} {\bibfnamefont {M.~B.}\ \bibnamefont {Maple}}, \bibinfo {author}
  {\bibfnamefont {J.}~\bibnamefont {Sarrao}}, \bibinfo {author} {\bibfnamefont
  {P.}~\bibnamefont {Armstrong}}, \bibinfo {author} {\bibfnamefont
  {Z.}~\bibnamefont {Fisk}},\ and\ \bibinfo {author} {\bibfnamefont
  {H.}~\bibnamefont {Yamagami}},\ }\href
  {https://doi.org/10.1016/S0368-2048(01)00257-2} {\bibfield  {journal}
  {\bibinfo  {journal} {Journal of Electron Spectroscopy and Related
  Phenomena}\ }\textbf {\bibinfo {volume} {117}},\ \bibinfo {pages} {347}
  (\bibinfo {year} {2001})}\BibitemShut {NoStop}%
\bibitem [{\citenamefont {Pei}\ \emph {et~al.}(2021)\citenamefont {Pei},
  \citenamefont {Zhang}, \citenamefont {Wei}, \citenamefont {Chen},
  \citenamefont {Hu}, \citenamefont {Yang}, \citenamefont {Yuan},\ and\
  \citenamefont {Qi}}]{Qi2021Ultrafast}%
  \BibitemOpen
  \bibfield  {author} {\bibinfo {author} {\bibfnamefont {Y.~H.}\ \bibnamefont
  {Pei}}, \bibinfo {author} {\bibfnamefont {Y.~J.}\ \bibnamefont {Zhang}},
  \bibinfo {author} {\bibfnamefont {Z.~X.}\ \bibnamefont {Wei}}, \bibinfo
  {author} {\bibfnamefont {Y.~X.}\ \bibnamefont {Chen}}, \bibinfo {author}
  {\bibfnamefont {K.}~\bibnamefont {Hu}}, \bibinfo {author} {\bibfnamefont
  {Y.~F.}\ \bibnamefont {Yang}}, \bibinfo {author} {\bibfnamefont {H.~Q.}\
  \bibnamefont {Yuan}},\ and\ \bibinfo {author} {\bibfnamefont
  {J.}~\bibnamefont {Qi}},\ }\href {https://arxiv.org/abs/2102.08572}
  {\bibfield  {journal} {\bibinfo  {journal} {arXiv preprint arXiv:2102.08572}\
  } (\bibinfo {year} {2021})}\BibitemShut {NoStop}%
\bibitem [{\citenamefont {Chen}\ \emph {et~al.}(2017)\citenamefont {Chen},
  \citenamefont {Xu}, \citenamefont {Niu}, \citenamefont {Jiang}, \citenamefont
  {Peng}, \citenamefont {Xu}, \citenamefont {Wen}, \citenamefont {Ding},
  \citenamefont {Huang}, \citenamefont {Shu}, \citenamefont {Zhang},
  \citenamefont {Lee}, \citenamefont {Strocov}, \citenamefont {Shi},
  \citenamefont {Bisti}, \citenamefont {Schmitt}, \citenamefont {Huang},
  \citenamefont {Dudin}, \citenamefont {Lai}, \citenamefont {Kirchner},
  \citenamefont {Yuan},\ and\ \citenamefont {Feng}}]{chen2017direct}%
  \BibitemOpen
  \bibfield  {author} {\bibinfo {author} {\bibfnamefont {Q.~Y.}\ \bibnamefont
  {Chen}}, \bibinfo {author} {\bibfnamefont {D.~F.}\ \bibnamefont {Xu}},
  \bibinfo {author} {\bibfnamefont {X.~H.}\ \bibnamefont {Niu}}, \bibinfo
  {author} {\bibfnamefont {J.}~\bibnamefont {Jiang}}, \bibinfo {author}
  {\bibfnamefont {R.}~\bibnamefont {Peng}}, \bibinfo {author} {\bibfnamefont
  {H.~C.}\ \bibnamefont {Xu}}, \bibinfo {author} {\bibfnamefont {H.~P.}\
  \bibnamefont {Wen}}, \bibinfo {author} {\bibfnamefont {Z.~F.}\ \bibnamefont
  {Ding}}, \bibinfo {author} {\bibfnamefont {K.}~\bibnamefont {Huang}},
  \bibinfo {author} {\bibfnamefont {L.}~\bibnamefont {Shu}}, \bibinfo {author}
  {\bibfnamefont {Y.~J.}\ \bibnamefont {Zhang}}, \bibinfo {author}
  {\bibfnamefont {H.}~\bibnamefont {Lee}}, \bibinfo {author} {\bibfnamefont
  {V.~N.}\ \bibnamefont {Strocov}}, \bibinfo {author} {\bibfnamefont
  {M.}~\bibnamefont {Shi}}, \bibinfo {author} {\bibfnamefont {F.}~\bibnamefont
  {Bisti}}, \bibinfo {author} {\bibfnamefont {T.}~\bibnamefont {Schmitt}},
  \bibinfo {author} {\bibfnamefont {Y.~B.}\ \bibnamefont {Huang}}, \bibinfo
  {author} {\bibfnamefont {P.}~\bibnamefont {Dudin}}, \bibinfo {author}
  {\bibfnamefont {X.~C.}\ \bibnamefont {Lai}}, \bibinfo {author} {\bibfnamefont
  {S.}~\bibnamefont {Kirchner}}, \bibinfo {author} {\bibfnamefont {H.~Q.}\
  \bibnamefont {Yuan}},\ and\ \bibinfo {author} {\bibfnamefont {D.~L.}\
  \bibnamefont {Feng}},\ }\href {https://doi.org/10.1103/PhysRevB.96.045107}
  {\bibfield  {journal} {\bibinfo  {journal} {Phys. Rev. B}\ }\textbf {\bibinfo
  {volume} {96}},\ \bibinfo {pages} {045107} (\bibinfo {year}
  {2017})}\BibitemShut {NoStop}%
\bibitem [{\citenamefont {Jang}\ \emph {et~al.}(2020)\citenamefont {Jang},
  \citenamefont {Denlinger}, \citenamefont {Allen}, \citenamefont {Zapf},
  \citenamefont {Maple}, \citenamefont {Kim}, \citenamefont {Jang},\ and\
  \citenamefont {Shim}}]{jang2017evolution}%
  \BibitemOpen
  \bibfield  {author} {\bibinfo {author} {\bibfnamefont {S.~Y.}\ \bibnamefont
  {Jang}}, \bibinfo {author} {\bibfnamefont {J.~D.}\ \bibnamefont {Denlinger}},
  \bibinfo {author} {\bibfnamefont {J.~W.}\ \bibnamefont {Allen}}, \bibinfo
  {author} {\bibfnamefont {V.~S.}\ \bibnamefont {Zapf}}, \bibinfo {author}
  {\bibfnamefont {M.~B.}\ \bibnamefont {Maple}}, \bibinfo {author}
  {\bibfnamefont {J.~N.}\ \bibnamefont {Kim}}, \bibinfo {author} {\bibfnamefont
  {B.~G.}\ \bibnamefont {Jang}},\ and\ \bibinfo {author} {\bibfnamefont
  {J.~H.}\ \bibnamefont {Shim}},\ }\href
  {https://doi.org/10.1073/pnas.2001778117} {\bibfield  {journal} {\bibinfo
  {journal} {Proc. Natl. Acad. Sci.}\ }\textbf {\bibinfo {volume} {117}},\
  \bibinfo {pages} {23467} (\bibinfo {year} {2020})}\BibitemShut {NoStop}%
\bibitem [{\citenamefont {Kroha}\ \emph {et~al.}(2003)\citenamefont {Kroha},
  \citenamefont {Kirchner}, \citenamefont {Sellier}, \citenamefont {W\"olfle},
  \citenamefont {Ehm}, \citenamefont {Reinert}, \citenamefont {H\"ufner},\ and\
  \citenamefont {Geibel}}]{kroha2003structure}%
  \BibitemOpen
  \bibfield  {author} {\bibinfo {author} {\bibfnamefont {J.}~\bibnamefont
  {Kroha}}, \bibinfo {author} {\bibfnamefont {S.}~\bibnamefont {Kirchner}},
  \bibinfo {author} {\bibfnamefont {G.}~\bibnamefont {Sellier}}, \bibinfo
  {author} {\bibfnamefont {P.}~\bibnamefont {W\"olfle}}, \bibinfo {author}
  {\bibfnamefont {D.}~\bibnamefont {Ehm}}, \bibinfo {author} {\bibfnamefont
  {F.}~\bibnamefont {Reinert}}, \bibinfo {author} {\bibfnamefont
  {S.}~\bibnamefont {H\"ufner}},\ and\ \bibinfo {author} {\bibfnamefont
  {C.}~\bibnamefont {Geibel}},\ }\href
  {https://doi.org/10.1016/S1386-9477(02)00977-3} {\bibfield  {journal}
  {\bibinfo  {journal} {Physica E: Low-dimensional Systems and Nanostructures}\
  }\textbf {\bibinfo {volume} {18}},\ \bibinfo {pages} {69} (\bibinfo {year}
  {2003})}\BibitemShut {NoStop}%
\bibitem [{\citenamefont {Kummer}\ \emph {et~al.}(2015)\citenamefont {Kummer},
  \citenamefont {Patil}, \citenamefont {Chikina}, \citenamefont {G\"uttler},
  \citenamefont {H\"oppner}, \citenamefont {Generalov}, \citenamefont
  {Danzenb\"acher}, \citenamefont {Seiro}, \citenamefont {Hannaske},
  \citenamefont {Krellner}, \citenamefont {Kucherenko}, \citenamefont {Shi},
  \citenamefont {Radovic}, \citenamefont {Rienks}, \citenamefont {Zwicknagl},
  \citenamefont {Matho}, \citenamefont {Allen}, \citenamefont {Laubschat},
  \citenamefont {Geibel},\ and\ \citenamefont
  {Vyalikh}}]{kummer2015temperature}%
  \BibitemOpen
  \bibfield  {author} {\bibinfo {author} {\bibfnamefont {K.}~\bibnamefont
  {Kummer}}, \bibinfo {author} {\bibfnamefont {S.}~\bibnamefont {Patil}},
  \bibinfo {author} {\bibfnamefont {A.}~\bibnamefont {Chikina}}, \bibinfo
  {author} {\bibfnamefont {M.}~\bibnamefont {G\"uttler}}, \bibinfo {author}
  {\bibfnamefont {M.}~\bibnamefont {H\"oppner}}, \bibinfo {author}
  {\bibfnamefont {A.}~\bibnamefont {Generalov}}, \bibinfo {author}
  {\bibfnamefont {S.}~\bibnamefont {Danzenb\"acher}}, \bibinfo {author}
  {\bibfnamefont {S.}~\bibnamefont {Seiro}}, \bibinfo {author} {\bibfnamefont
  {A.}~\bibnamefont {Hannaske}}, \bibinfo {author} {\bibfnamefont
  {C.}~\bibnamefont {Krellner}}, \bibinfo {author} {\bibfnamefont
  {Y.}~\bibnamefont {Kucherenko}}, \bibinfo {author} {\bibfnamefont
  {M.}~\bibnamefont {Shi}}, \bibinfo {author} {\bibfnamefont {M.}~\bibnamefont
  {Radovic}}, \bibinfo {author} {\bibfnamefont {E.}~\bibnamefont {Rienks}},
  \bibinfo {author} {\bibfnamefont {G.}~\bibnamefont {Zwicknagl}}, \bibinfo
  {author} {\bibfnamefont {K.}~\bibnamefont {Matho}}, \bibinfo {author}
  {\bibfnamefont {J.~W.}\ \bibnamefont {Allen}}, \bibinfo {author}
  {\bibfnamefont {C.}~\bibnamefont {Laubschat}}, \bibinfo {author}
  {\bibfnamefont {C.}~\bibnamefont {Geibel}},\ and\ \bibinfo {author}
  {\bibfnamefont {D.~V.}\ \bibnamefont {Vyalikh}},\ }\href
  {https://doi.org/10.1103/PhysRevX.5.011028} {\bibfield  {journal} {\bibinfo
  {journal} {Phys. Rev. X}\ }\textbf {\bibinfo {volume} {5}},\ \bibinfo {pages}
  {011028} (\bibinfo {year} {2015})}\BibitemShut {NoStop}%
\bibitem [{\citenamefont {Shu}\ \emph {et~al.}(2021)\citenamefont {Shu},
  \citenamefont {Adroja}, \citenamefont {Hillier}, \citenamefont {Zhang},
  \citenamefont {Chen}, \citenamefont {Shen}, \citenamefont {Orlandi},
  \citenamefont {Walker}, \citenamefont {Liu}, \citenamefont {Cao},
  \citenamefont {Steglich}, \citenamefont {Yuan},\ and\ \citenamefont
  {Smidman}}]{micheal2020}%
  \BibitemOpen
  \bibfield  {author} {\bibinfo {author} {\bibfnamefont {J.~W.}\ \bibnamefont
  {Shu}}, \bibinfo {author} {\bibfnamefont {D.~T.}\ \bibnamefont {Adroja}},
  \bibinfo {author} {\bibfnamefont {A.~D.}\ \bibnamefont {Hillier}}, \bibinfo
  {author} {\bibfnamefont {Y.~J.}\ \bibnamefont {Zhang}}, \bibinfo {author}
  {\bibfnamefont {Y.~X.}\ \bibnamefont {Chen}}, \bibinfo {author}
  {\bibfnamefont {B.}~\bibnamefont {Shen}}, \bibinfo {author} {\bibfnamefont
  {F.}~\bibnamefont {Orlandi}}, \bibinfo {author} {\bibfnamefont {H.~C.}\
  \bibnamefont {Walker}}, \bibinfo {author} {\bibfnamefont {Y.}~\bibnamefont
  {Liu}}, \bibinfo {author} {\bibfnamefont {C.}~\bibnamefont {Cao}}, \bibinfo
  {author} {\bibfnamefont {F.}~\bibnamefont {Steglich}}, \bibinfo {author}
  {\bibfnamefont {H.~Q.}\ \bibnamefont {Yuan}},\ and\ \bibinfo {author}
  {\bibfnamefont {M.}~\bibnamefont {Smidman}},\ }\href
  {https://arxiv.org/abs/2102.12788} {\bibfield  {journal} {\bibinfo  {journal}
  {arXiv preprint arXiv:2102.12788}\ } (\bibinfo {year} {2021})}\BibitemShut
  {NoStop}%
\bibitem [{\citenamefont {Ehm}\ \emph {et~al.}(2007)\citenamefont {Ehm},
  \citenamefont {H\"ufner}, \citenamefont {Reinert}, \citenamefont {Kroha},
  \citenamefont {W\"lfle}, \citenamefont {Stockert}, \citenamefont {Geibel},\
  and\ \citenamefont {L\"ohneysen}}]{ehm2007high}%
  \BibitemOpen
  \bibfield  {author} {\bibinfo {author} {\bibfnamefont {D.}~\bibnamefont
  {Ehm}}, \bibinfo {author} {\bibfnamefont {S.}~\bibnamefont {H\"ufner}},
  \bibinfo {author} {\bibfnamefont {F.}~\bibnamefont {Reinert}}, \bibinfo
  {author} {\bibfnamefont {J.}~\bibnamefont {Kroha}}, \bibinfo {author}
  {\bibfnamefont {P.}~\bibnamefont {W\"lfle}}, \bibinfo {author} {\bibfnamefont
  {O.}~\bibnamefont {Stockert}}, \bibinfo {author} {\bibfnamefont
  {C.}~\bibnamefont {Geibel}},\ and\ \bibinfo {author} {\bibfnamefont {H.~v.}\
  \bibnamefont {L\"ohneysen}},\ }\href
  {https://doi.org/10.1103/PhysRevB.76.045117} {\bibfield  {journal} {\bibinfo
  {journal} {Phys. Rev. B}\ }\textbf {\bibinfo {volume} {76}},\ \bibinfo
  {pages} {045117} (\bibinfo {year} {2007})}\BibitemShut {NoStop}%
\bibitem [{\citenamefont {Im}\ \emph {et~al.}(2008)\citenamefont {Im},
  \citenamefont {Ito}, \citenamefont {Kim}, \citenamefont {Kimura},
  \citenamefont {Lee}, \citenamefont {Hong}, \citenamefont {Kwon},
  \citenamefont {Yasui},\ and\ \citenamefont {Yamagami}}]{im2008direct}%
  \BibitemOpen
  \bibfield  {author} {\bibinfo {author} {\bibfnamefont {H.~J.}\ \bibnamefont
  {Im}}, \bibinfo {author} {\bibfnamefont {T.}~\bibnamefont {Ito}}, \bibinfo
  {author} {\bibfnamefont {H.-D.}\ \bibnamefont {Kim}}, \bibinfo {author}
  {\bibfnamefont {S.}~\bibnamefont {Kimura}}, \bibinfo {author} {\bibfnamefont
  {K.~E.}\ \bibnamefont {Lee}}, \bibinfo {author} {\bibfnamefont {J.~B.}\
  \bibnamefont {Hong}}, \bibinfo {author} {\bibfnamefont {Y.~S.}\ \bibnamefont
  {Kwon}}, \bibinfo {author} {\bibfnamefont {A.}~\bibnamefont {Yasui}},\ and\
  \bibinfo {author} {\bibfnamefont {H.}~\bibnamefont {Yamagami}},\ }\href
  {https://doi.org/10.1103/PhysRevLett.100.176402} {\bibfield  {journal}
  {\bibinfo  {journal} {Phys. Rev. Lett.}\ }\textbf {\bibinfo {volume} {100}},\
  \bibinfo {pages} {176402} (\bibinfo {year} {2008})}\BibitemShut {NoStop}%
\bibitem [{\citenamefont {Coleman}(2007)}]{HF2007coleman}%
  \BibitemOpen
  \bibfield  {author} {\bibinfo {author} {\bibfnamefont {P.}~\bibnamefont
  {Coleman}},\ }\href@noop {} {\emph {\bibinfo {title} {Heavy Fermions:
  electrons at the edge of magnetism. Handbook of Magnetism and Advanced
  Magnetic Materials}}},\ Vol.~\bibinfo {volume} {1}\ (\bibinfo  {publisher}
  {Wiley, New York},\ \bibinfo {year} {2007})\BibitemShut {NoStop}%
\bibitem [{\citenamefont {Komijani}\ and\ \citenamefont
  {Coleman}(2018)}]{Komijani2018Model}%
  \BibitemOpen
  \bibfield  {author} {\bibinfo {author} {\bibfnamefont {Y.}~\bibnamefont
  {Komijani}}\ and\ \bibinfo {author} {\bibfnamefont {P.}~\bibnamefont
  {Coleman}},\ }\href {https://doi.org/10.1103/PhysRevLett.120.157206}
  {\bibfield  {journal} {\bibinfo  {journal} {Phys. Rev. Lett.}\ }\textbf
  {\bibinfo {volume} {120}},\ \bibinfo {pages} {157206} (\bibinfo {year}
  {2018})}\BibitemShut {NoStop}%
\bibitem [{\citenamefont {Kirkpatrick}\ and\ \citenamefont
  {Belitz}(2020)}]{kirkpatrick2020ferromagnetic}%
  \BibitemOpen
  \bibfield  {author} {\bibinfo {author} {\bibfnamefont {T.~R.}\ \bibnamefont
  {Kirkpatrick}}\ and\ \bibinfo {author} {\bibfnamefont {D.}~\bibnamefont
  {Belitz}},\ }\href {https://doi.org/10.1103/PhysRevLett.124.147201}
  {\bibfield  {journal} {\bibinfo  {journal} {Phys. Rev. Lett.}\ }\textbf
  {\bibinfo {volume} {124}},\ \bibinfo {pages} {147201} (\bibinfo {year}
  {2020})}\BibitemShut {NoStop}%
\bibitem [{\citenamefont {Yano}\ \emph {et~al.}(2007)\citenamefont {Yano},
  \citenamefont {Sekiyama}, \citenamefont {Fujiwara}, \citenamefont {Saita},
  \citenamefont {Imada}, \citenamefont {Muro}, \citenamefont {Onuki},\ and\
  \citenamefont {Suga}}]{Yano2007Three}%
  \BibitemOpen
  \bibfield  {author} {\bibinfo {author} {\bibfnamefont {M.}~\bibnamefont
  {Yano}}, \bibinfo {author} {\bibfnamefont {A.}~\bibnamefont {Sekiyama}},
  \bibinfo {author} {\bibfnamefont {A.}~\bibnamefont {Fujiwara}}, \bibinfo
  {author} {\bibfnamefont {T.}~\bibnamefont {Saita}}, \bibinfo {author}
  {\bibfnamefont {S.}~\bibnamefont {Imada}}, \bibinfo {author} {\bibfnamefont
  {T.}~\bibnamefont {Muro}}, \bibinfo {author} {\bibfnamefont {Y.}~\bibnamefont
  {Onuki}},\ and\ \bibinfo {author} {\bibfnamefont {S.}~\bibnamefont {Suga}},\
  }\href {https://doi.org/10.1103/PhysRevLett.98.036405} {\bibfield  {journal}
  {\bibinfo  {journal} {Phys. Rev. Lett.}\ }\textbf {\bibinfo {volume} {98}},\
  \bibinfo {pages} {036405} (\bibinfo {year} {2007})}\BibitemShut {NoStop}%
\bibitem [{\citenamefont {Generalov}\ \emph {et~al.}(2017)\citenamefont
  {Generalov}, \citenamefont {Sokolov}, \citenamefont {Chikina}, \citenamefont
  {Kucherenko}, \citenamefont {Antonov}, \citenamefont {Bekenov}, \citenamefont
  {Patil}, \citenamefont {Huxley}, \citenamefont {Allen}, \citenamefont
  {Matho}, \citenamefont {Kummer}, \citenamefont {Vyalikh},\ and\ \citenamefont
  {Laubschat}}]{generalov2017insight}%
  \BibitemOpen
  \bibfield  {author} {\bibinfo {author} {\bibfnamefont {A.}~\bibnamefont
  {Generalov}}, \bibinfo {author} {\bibfnamefont {D.~A.}\ \bibnamefont
  {Sokolov}}, \bibinfo {author} {\bibfnamefont {A.}~\bibnamefont {Chikina}},
  \bibinfo {author} {\bibfnamefont {Y.}~\bibnamefont {Kucherenko}}, \bibinfo
  {author} {\bibfnamefont {V.~N.}\ \bibnamefont {Antonov}}, \bibinfo {author}
  {\bibfnamefont {L.~V.}\ \bibnamefont {Bekenov}}, \bibinfo {author}
  {\bibfnamefont {S.}~\bibnamefont {Patil}}, \bibinfo {author} {\bibfnamefont
  {A.~D.}\ \bibnamefont {Huxley}}, \bibinfo {author} {\bibfnamefont {J.~W.}\
  \bibnamefont {Allen}}, \bibinfo {author} {\bibfnamefont {K.}~\bibnamefont
  {Matho}}, \bibinfo {author} {\bibfnamefont {K.}~\bibnamefont {Kummer}},
  \bibinfo {author} {\bibfnamefont {D.~V.}\ \bibnamefont {Vyalikh}},\ and\
  \bibinfo {author} {\bibfnamefont {C.}~\bibnamefont {Laubschat}},\ }\href
  {https://doi.org/10.1103/PhysRevB.95.184433} {\bibfield  {journal} {\bibinfo
  {journal} {Phys. Rev. B}\ }\textbf {\bibinfo {volume} {95}},\ \bibinfo
  {pages} {184433} (\bibinfo {year} {2017})}\BibitemShut {NoStop}%
\bibitem [{\citenamefont {Yamaoka}\ \emph {et~al.}(2017)\citenamefont
  {Yamaoka}, \citenamefont {Thunstr\"om}, \citenamefont {Tsujii}, \citenamefont
  {Katoh}, \citenamefont {Yamamoto}, \citenamefont {Schwier}, \citenamefont
  {Shimada}, \citenamefont {Iwasawa}, \citenamefont {Arita}, \citenamefont
  {Jarrige}, \citenamefont {Hiraoka}, \citenamefont {Ishii}, \citenamefont
  {Tsuei},\ and\ \citenamefont {Mizuki}}]{yamaoka2017electronic}%
  \BibitemOpen
  \bibfield  {author} {\bibinfo {author} {\bibfnamefont {H.}~\bibnamefont
  {Yamaoka}}, \bibinfo {author} {\bibfnamefont {P.}~\bibnamefont
  {Thunstr\"om}}, \bibinfo {author} {\bibfnamefont {N.}~\bibnamefont {Tsujii}},
  \bibinfo {author} {\bibfnamefont {K.}~\bibnamefont {Katoh}}, \bibinfo
  {author} {\bibfnamefont {Y.}~\bibnamefont {Yamamoto}}, \bibinfo {author}
  {\bibfnamefont {E.~F.}\ \bibnamefont {Schwier}}, \bibinfo {author}
  {\bibfnamefont {K.}~\bibnamefont {Shimada}}, \bibinfo {author} {\bibfnamefont
  {H.}~\bibnamefont {Iwasawa}}, \bibinfo {author} {\bibfnamefont
  {M.}~\bibnamefont {Arita}}, \bibinfo {author} {\bibfnamefont
  {I.}~\bibnamefont {Jarrige}}, \bibinfo {author} {\bibfnamefont
  {N.}~\bibnamefont {Hiraoka}}, \bibinfo {author} {\bibfnamefont
  {H.}~\bibnamefont {Ishii}}, \bibinfo {author} {\bibfnamefont {K.-D.}\
  \bibnamefont {Tsuei}},\ and\ \bibinfo {author} {\bibfnamefont
  {J.}~\bibnamefont {Mizuki}},\ }\href
  {https://doi.org/10.1088/1361-648X/aa8b98} {\bibfield  {journal} {\bibinfo
  {journal} {Journal of Physics: Condensed Matter}\ }\textbf {\bibinfo {volume}
  {29}},\ \bibinfo {pages} {475502} (\bibinfo {year} {2017})}\BibitemShut
  {NoStop}%
\bibitem [{\citenamefont {Hafner}\ \emph {et~al.}(2019)\citenamefont {Hafner},
  \citenamefont {Rai}, \citenamefont {Banda}, \citenamefont {Kliemt},
  \citenamefont {Krellner}, \citenamefont {Sichelschmidt}, \citenamefont
  {Morosan}, \citenamefont {Geibel},\ and\ \citenamefont
  {Brando}}]{hafner2019kondo}%
  \BibitemOpen
  \bibfield  {author} {\bibinfo {author} {\bibfnamefont {D.}~\bibnamefont
  {Hafner}}, \bibinfo {author} {\bibfnamefont {B.~K.}\ \bibnamefont {Rai}},
  \bibinfo {author} {\bibfnamefont {J.}~\bibnamefont {Banda}}, \bibinfo
  {author} {\bibfnamefont {K.}~\bibnamefont {Kliemt}}, \bibinfo {author}
  {\bibfnamefont {C.}~\bibnamefont {Krellner}}, \bibinfo {author}
  {\bibfnamefont {J.}~\bibnamefont {Sichelschmidt}}, \bibinfo {author}
  {\bibfnamefont {E.}~\bibnamefont {Morosan}}, \bibinfo {author} {\bibfnamefont
  {C.}~\bibnamefont {Geibel}},\ and\ \bibinfo {author} {\bibfnamefont
  {M.}~\bibnamefont {Brando}},\ }\href
  {https://doi.org/10.1103/PhysRevB.99.201109} {\bibfield  {journal} {\bibinfo
  {journal} {Phys. Rev. B}\ }\textbf {\bibinfo {volume} {99}},\ \bibinfo
  {pages} {201109} (\bibinfo {year} {2019})}\BibitemShut {NoStop}%
\bibitem [{\citenamefont {Ahamed}\ \emph {et~al.}(2018)\citenamefont {Ahamed},
  \citenamefont {Moessner},\ and\ \citenamefont {Erten}}]{Ahamed2018PRB}%
  \BibitemOpen
  \bibfield  {author} {\bibinfo {author} {\bibfnamefont {S.}~\bibnamefont
  {Ahamed}}, \bibinfo {author} {\bibfnamefont {R.}~\bibnamefont {Moessner}},\
  and\ \bibinfo {author} {\bibfnamefont {O.}~\bibnamefont {Erten}},\ }\href
  {https://doi.org/10.1103/PhysRevB.98.054420} {\bibfield  {journal} {\bibinfo
  {journal} {Phys. Rev. B}\ }\textbf {\bibinfo {volume} {98}},\ \bibinfo
  {pages} {054420} (\bibinfo {year} {2018})}\BibitemShut {NoStop}%
\bibitem [{\citenamefont {Li}\ \emph {et~al.}(2010)\citenamefont {Li},
  \citenamefont {Zhang},\ and\ \citenamefont {Yu}}]{Li2010PRB}%
  \BibitemOpen
  \bibfield  {author} {\bibinfo {author} {\bibfnamefont {G.~B.}\ \bibnamefont
  {Li}}, \bibinfo {author} {\bibfnamefont {G.~M.}\ \bibnamefont {Zhang}},\ and\
  \bibinfo {author} {\bibfnamefont {L.}~\bibnamefont {Yu}},\ }\href
  {https://doi.org/10.1103/PhysRevB.81.094420} {\bibfield  {journal} {\bibinfo
  {journal} {Phys. Rev. B}\ }\textbf {\bibinfo {volume} {81}},\ \bibinfo
  {pages} {094420} (\bibinfo {year} {2010})}\BibitemShut {NoStop}%
\bibitem [{\citenamefont {Bernhard}\ and\ \citenamefont
  {Lacroix}(2015)}]{Bernhard2015PRB}%
  \BibitemOpen
  \bibfield  {author} {\bibinfo {author} {\bibfnamefont {B.}~\bibnamefont
  {Bernhard}}\ and\ \bibinfo {author} {\bibfnamefont {C.}~\bibnamefont
  {Lacroix}},\ }\href {https://doi.org/10.1103/PhysRevB.92.094401} {\bibfield
  {journal} {\bibinfo  {journal} {Phys. Rev. B}\ }\textbf {\bibinfo {volume}
  {92}},\ \bibinfo {pages} {094401} (\bibinfo {year} {2015})}\BibitemShut
  {NoStop}%
\bibitem [{\citenamefont {Smidman}\ \emph {et~al.}(2018)\citenamefont
  {Smidman}, \citenamefont {Stockert}, \citenamefont {Arndt}, \citenamefont
  {Pang}, \citenamefont {Jiao}, \citenamefont {Yuan}, \citenamefont {Vieyra},
  \citenamefont {Kitagawa}, \citenamefont {Ishida}, \citenamefont {Fujiwara},
  \citenamefont {Kobayashi}, \citenamefont {Schuberth}, \citenamefont
  {Tippmann}, \citenamefont {Steinke}, \citenamefont {Lausberg}, \citenamefont
  {Steppke}, \citenamefont {Brando}, \citenamefont {Pfau}, \citenamefont
  {Stockert}, \citenamefont {Sun}, \citenamefont {Friedemann}, \citenamefont
  {Wirth}, \citenamefont {Krellner}, \citenamefont {Kirchner}, \citenamefont
  {Nica}, \citenamefont {Yu}, \citenamefont {Si},\ and\ \citenamefont
  {Steglich}}]{smidman2018interplay}%
  \BibitemOpen
  \bibfield  {author} {\bibinfo {author} {\bibfnamefont {M.}~\bibnamefont
  {Smidman}}, \bibinfo {author} {\bibfnamefont {O.}~\bibnamefont {Stockert}},
  \bibinfo {author} {\bibfnamefont {J.}~\bibnamefont {Arndt}}, \bibinfo
  {author} {\bibfnamefont {G.}~\bibnamefont {Pang}}, \bibinfo {author}
  {\bibfnamefont {L.}~\bibnamefont {Jiao}}, \bibinfo {author} {\bibfnamefont
  {H.~Q.}\ \bibnamefont {Yuan}}, \bibinfo {author} {\bibfnamefont {H.~A.}\
  \bibnamefont {Vieyra}}, \bibinfo {author} {\bibfnamefont {S.}~\bibnamefont
  {Kitagawa}}, \bibinfo {author} {\bibfnamefont {K.}~\bibnamefont {Ishida}},
  \bibinfo {author} {\bibfnamefont {K.}~\bibnamefont {Fujiwara}}, \bibinfo
  {author} {\bibfnamefont {T.~C.}\ \bibnamefont {Kobayashi}}, \bibinfo {author}
  {\bibfnamefont {E.}~\bibnamefont {Schuberth}}, \bibinfo {author}
  {\bibfnamefont {M.}~\bibnamefont {Tippmann}}, \bibinfo {author}
  {\bibfnamefont {L.}~\bibnamefont {Steinke}}, \bibinfo {author} {\bibfnamefont
  {S.}~\bibnamefont {Lausberg}}, \bibinfo {author} {\bibfnamefont
  {A.}~\bibnamefont {Steppke}}, \bibinfo {author} {\bibfnamefont
  {M.}~\bibnamefont {Brando}}, \bibinfo {author} {\bibfnamefont
  {H.}~\bibnamefont {Pfau}}, \bibinfo {author} {\bibfnamefont {U.}~\bibnamefont
  {Stockert}}, \bibinfo {author} {\bibfnamefont {P.}~\bibnamefont {Sun}},
  \bibinfo {author} {\bibfnamefont {S.}~\bibnamefont {Friedemann}}, \bibinfo
  {author} {\bibfnamefont {S.}~\bibnamefont {Wirth}}, \bibinfo {author}
  {\bibfnamefont {C.}~\bibnamefont {Krellner}}, \bibinfo {author}
  {\bibfnamefont {S.}~\bibnamefont {Kirchner}}, \bibinfo {author}
  {\bibfnamefont {E.~M.}\ \bibnamefont {Nica}}, \bibinfo {author}
  {\bibfnamefont {R.}~\bibnamefont {Yu}}, \bibinfo {author} {\bibfnamefont
  {Q.}~\bibnamefont {Si}},\ and\ \bibinfo {author} {\bibfnamefont
  {F.}~\bibnamefont {Steglich}},\ }\href
  {https://doi.org/10.1080/14786435.2018.1511070} {\bibfield  {journal}
  {\bibinfo  {journal} {Philosophical Magazine}\ }\textbf {\bibinfo {volume}
  {98}},\ \bibinfo {pages} {2930} (\bibinfo {year} {2018})}\BibitemShut
  {NoStop}%
\bibitem [{\citenamefont {Wang}\ \emph {et~al.}(2021)\citenamefont {Wang},
  \citenamefont {Du}, \citenamefont {Zhang}, \citenamefont {Graf},
  \citenamefont {Shen}, \citenamefont {Chen}, \citenamefont {Liu},
  \citenamefont {Smidman}, \citenamefont {Cao}, \citenamefont {Steglich},\ and\
  \citenamefont {Yuan}}]{An2020QO}%
  \BibitemOpen
  \bibfield  {author} {\bibinfo {author} {\bibfnamefont {A.}~\bibnamefont
  {Wang}}, \bibinfo {author} {\bibfnamefont {F.}~\bibnamefont {Du}}, \bibinfo
  {author} {\bibfnamefont {Y.~J.}\ \bibnamefont {Zhang}}, \bibinfo {author}
  {\bibfnamefont {D.}~\bibnamefont {Graf}}, \bibinfo {author} {\bibfnamefont
  {B.}~\bibnamefont {Shen}}, \bibinfo {author} {\bibfnamefont {Y.}~\bibnamefont
  {Chen}}, \bibinfo {author} {\bibfnamefont {Y.}~\bibnamefont {Liu}}, \bibinfo
  {author} {\bibfnamefont {M.}~\bibnamefont {Smidman}}, \bibinfo {author}
  {\bibfnamefont {C.}~\bibnamefont {Cao}}, \bibinfo {author} {\bibfnamefont
  {F.}~\bibnamefont {Steglich}},\ and\ \bibinfo {author} {\bibfnamefont
  {H.~Q.}\ \bibnamefont {Yuan}},\ }\href {https://arxiv.org/abs/2101.08972}
  {\bibfield  {journal} {\bibinfo  {journal} {arXiv preprint arXiv:2101.08972}\
  } (\bibinfo {year} {2021})}\BibitemShut {NoStop}%
\bibitem [{\citenamefont {Shishido}\ \emph {et~al.}(2002)\citenamefont
  {Shishido}, \citenamefont {Settai}, \citenamefont {Aoki}, \citenamefont
  {Ikeda}, \citenamefont {Nakawaki}, \citenamefont {Nakamura}, \citenamefont
  {Iizuka}, \citenamefont {Inada}, \citenamefont {Sugiyama}, \citenamefont
  {Takeuchi}, \citenamefont {Kindo}, \citenamefont {Kobayashi}, \citenamefont
  {Haga}, \citenamefont {Harima}, \citenamefont {Aoki}, \citenamefont {Namiki},
  \citenamefont {Sato},\ and\ \citenamefont {\={O}nuki}}]{shishido2002JPSJ}%
  \BibitemOpen
  \bibfield  {author} {\bibinfo {author} {\bibfnamefont {H.}~\bibnamefont
  {Shishido}}, \bibinfo {author} {\bibfnamefont {R.}~\bibnamefont {Settai}},
  \bibinfo {author} {\bibfnamefont {D.}~\bibnamefont {Aoki}}, \bibinfo {author}
  {\bibfnamefont {S.}~\bibnamefont {Ikeda}}, \bibinfo {author} {\bibfnamefont
  {H.}~\bibnamefont {Nakawaki}}, \bibinfo {author} {\bibfnamefont
  {N.}~\bibnamefont {Nakamura}}, \bibinfo {author} {\bibfnamefont
  {T.}~\bibnamefont {Iizuka}}, \bibinfo {author} {\bibfnamefont
  {Y.}~\bibnamefont {Inada}}, \bibinfo {author} {\bibfnamefont
  {K.}~\bibnamefont {Sugiyama}}, \bibinfo {author} {\bibfnamefont
  {T.}~\bibnamefont {Takeuchi}}, \bibinfo {author} {\bibfnamefont
  {K.}~\bibnamefont {Kindo}}, \bibinfo {author} {\bibfnamefont {T.~C.}\
  \bibnamefont {Kobayashi}}, \bibinfo {author} {\bibfnamefont {Y.}~\bibnamefont
  {Haga}}, \bibinfo {author} {\bibfnamefont {H.}~\bibnamefont {Harima}},
  \bibinfo {author} {\bibfnamefont {Y.}~\bibnamefont {Aoki}}, \bibinfo {author}
  {\bibfnamefont {T.}~\bibnamefont {Namiki}}, \bibinfo {author} {\bibfnamefont
  {H.}~\bibnamefont {Sato}},\ and\ \bibinfo {author} {\bibfnamefont
  {Y.}~\bibnamefont {\={O}nuki}},\ }\href {https://doi.org/10.1143/JPSJ.71.162}
  {\bibfield  {journal} {\bibinfo  {journal} {Journal of the Physical Society
  of Japan}\ }\textbf {\bibinfo {volume} {71}},\ \bibinfo {pages} {162}
  (\bibinfo {year} {2002})}\BibitemShut {NoStop}%
\bibitem [{\citenamefont {Harrison}\ \emph {et~al.}(2004)\citenamefont
  {Harrison}, \citenamefont {Alver}, \citenamefont {Goodrich}, \citenamefont
  {Vekhter}, \citenamefont {Sarrao}, \citenamefont {Pagliuso}, \citenamefont
  {Moreno}, \citenamefont {Balicas}, \citenamefont {Fisk}, \citenamefont
  {Hall}, \citenamefont {Macaluso},\ and\ \citenamefont
  {Chan}}]{Harrison2004PRL}%
  \BibitemOpen
  \bibfield  {author} {\bibinfo {author} {\bibfnamefont {N.}~\bibnamefont
  {Harrison}}, \bibinfo {author} {\bibfnamefont {U.}~\bibnamefont {Alver}},
  \bibinfo {author} {\bibfnamefont {R.~G.}\ \bibnamefont {Goodrich}}, \bibinfo
  {author} {\bibfnamefont {I.}~\bibnamefont {Vekhter}}, \bibinfo {author}
  {\bibfnamefont {J.~L.}\ \bibnamefont {Sarrao}}, \bibinfo {author}
  {\bibfnamefont {P.~G.}\ \bibnamefont {Pagliuso}}, \bibinfo {author}
  {\bibfnamefont {N.~O.}\ \bibnamefont {Moreno}}, \bibinfo {author}
  {\bibfnamefont {L.}~\bibnamefont {Balicas}}, \bibinfo {author} {\bibfnamefont
  {Z.}~\bibnamefont {Fisk}}, \bibinfo {author} {\bibfnamefont {D.}~\bibnamefont
  {Hall}}, \bibinfo {author} {\bibfnamefont {R.~T.}\ \bibnamefont {Macaluso}},\
  and\ \bibinfo {author} {\bibfnamefont {J.~Y.}\ \bibnamefont {Chan}},\ }\href
  {https://doi.org/10.1103/PhysRevLett.93.186405} {\bibfield  {journal}
  {\bibinfo  {journal} {Phys. Rev. Lett.}\ }\textbf {\bibinfo {volume} {93}},\
  \bibinfo {pages} {186405} (\bibinfo {year} {2004})}\BibitemShut {NoStop}%
\bibitem [{\citenamefont {Chen}\ \emph {et~al.}(2018)\citenamefont {Chen},
  \citenamefont {Xu}, \citenamefont {Niu}, \citenamefont {Peng}, \citenamefont
  {Xu}, \citenamefont {Wen}, \citenamefont {Liu}, \citenamefont {Shu},
  \citenamefont {Tan}, \citenamefont {Lai}, \citenamefont {Zhang},
  \citenamefont {Lee}, \citenamefont {Strocov}, \citenamefont {Bisti},
  \citenamefont {Dudin}, \citenamefont {Zhu}, \citenamefont {Yuan},
  \citenamefont {Kirchner},\ and\ \citenamefont {Feng}}]{chen2018band}%
  \BibitemOpen
  \bibfield  {author} {\bibinfo {author} {\bibfnamefont {Q.~Y.}\ \bibnamefont
  {Chen}}, \bibinfo {author} {\bibfnamefont {D.~F.}\ \bibnamefont {Xu}},
  \bibinfo {author} {\bibfnamefont {X.~H.}\ \bibnamefont {Niu}}, \bibinfo
  {author} {\bibfnamefont {R.}~\bibnamefont {Peng}}, \bibinfo {author}
  {\bibfnamefont {H.~C.}\ \bibnamefont {Xu}}, \bibinfo {author} {\bibfnamefont
  {C.~H.~P.}\ \bibnamefont {Wen}}, \bibinfo {author} {\bibfnamefont
  {X.}~\bibnamefont {Liu}}, \bibinfo {author} {\bibfnamefont {L.}~\bibnamefont
  {Shu}}, \bibinfo {author} {\bibfnamefont {S.~Y.}\ \bibnamefont {Tan}},
  \bibinfo {author} {\bibfnamefont {X.~C.}\ \bibnamefont {Lai}}, \bibinfo
  {author} {\bibfnamefont {Y.~J.}\ \bibnamefont {Zhang}}, \bibinfo {author}
  {\bibfnamefont {H.}~\bibnamefont {Lee}}, \bibinfo {author} {\bibfnamefont
  {V.~N.}\ \bibnamefont {Strocov}}, \bibinfo {author} {\bibfnamefont
  {F.}~\bibnamefont {Bisti}}, \bibinfo {author} {\bibfnamefont
  {P.}~\bibnamefont {Dudin}}, \bibinfo {author} {\bibfnamefont {J.~X.}\
  \bibnamefont {Zhu}}, \bibinfo {author} {\bibfnamefont {H.~Q.}\ \bibnamefont
  {Yuan}}, \bibinfo {author} {\bibfnamefont {S.}~\bibnamefont {Kirchner}},\
  and\ \bibinfo {author} {\bibfnamefont {D.~L.}\ \bibnamefont {Feng}},\ }\href
  {https://doi.org/10.1103/PhysRevLett.120.066403} {\bibfield  {journal}
  {\bibinfo  {journal} {Phys. Rev. Lett.}\ }\textbf {\bibinfo {volume} {120}},\
  \bibinfo {pages} {066403} (\bibinfo {year} {2018})}\BibitemShut {NoStop}%
\end{thebibliography}
\end{document}